\documentclass[aps,onecolumn,nofootinbib]{revtex4}

\usepackage{graphicx}% Include figure files
\usepackage{dcolumn}% Align table columns on decimal point
\usepackage{bm}% bold math
\usepackage{array}
\usepackage{booktabs}
\usepackage{multirow}
\newcommand{\head}[1]{\textnormal{\textbf{#1}}}
\newcommand{\normal}[1]{\multicolumn{1}{l}{#1}}
\usepackage{amssymb,amsfonts}
\usepackage{epsfig}
\usepackage{epstopdf}
\usepackage[all]{xy}
\usepackage{amsthm}
\usepackage{dcolumn}
\usepackage{hyperref}
\usepackage{url}
\usepackage{dsfont}
\usepackage{slashed}
\usepackage{mathrsfs}
\usepackage{lipsum}

\newcommand{\be}{\begin{equation}}
\newcommand{\ee}{\end{equation}}
\newcommand{\bea}{\begin{eqnarray}}

\newcommand{\eea}{\end{eqnarray}}

\def\inbar{\,\vrule height1.5ex width.4pt depth0pt}
\def\IR{\relax{\rm I\kern-.18em R}}
\def\IC{\relax\hbox{$\inbar\kern-.3em{\rm C}$}}

\begin{document}

\title{Nonlinear corrections to the single differential cross section for neutral current $e^{-} p$ scattering at the NLO approximation}

\author{S. Zarrin\footnote{zarrin@phys.usb.ac.ir}}

\author{S. Dadfar}

\affiliation{ Department of Physics, University of Sistan and Baluchestan, Zahedan, Iran}

\begin{abstract}
We present the effects of nonlinear corrections to the single differential cross section $d\sigma/dQ^{2}$ and the reduced cross section $\sigma_{r}(x,Q^{2})$ for the neutral current (NC) $e^{-} p$ scattering at the leading order (LO) and the next-to-leading order (NLO) approximations in perturbative quantum chromodynamics (QCD). Technically,  based on the double Laplace transform method, we first derive  the effects of the nonlinear corrections to the proton structure functions $F_{2}(x,Q^{2})$ and $F_{L}(x,Q^{2})$ and consequently obtain the corresponding single differential and reduced cross sections. Our results clearly indicate the consistency of the nonlinear behavior of the quark and gluon distributions at low $x$ values. Our numerical results ( obtained in a range of the virtuality $8.5<Q^{2}<5000$ $GeV^{2}$ and the Bjorken scale $10^{-5}<x<1$)  show that the effects of these nonlinear corrections to the proton structure functions are more noticeable at $x<0.001$ and, to some extent, control the incremental trend of these functions at low $x$ values.  Moreover, a comparison of our numerical results of the single differential and reduced cross sections at the NLO approximations with those of H1 collaboration data shows that the nonlinear corrections increase the accuracy of calculations rather than the linear calculations at low to moderate $Q^{2}$ values for low $x$ values.
\newline

PACS numbers: 14.20.Dh, 12.38.Bx, 13.60.Hb, 12.39.-x
 
\end{abstract}
\maketitle
%\tableofcontents
\section*{I. INTRODUCTION}
The measurements of inclusive deep inelastic scattering are important and fundamental to understanding the substructure of the proton. Within the framework of perturbative quantum chromodynamics (pQCD), the parton distribution functions (PDF's) describe the substructure of the proton.  These functions at a starting scale cannot be predicted by this framework and must be determined by fits to data using ad-hoc parameterisations \cite{1}. But, pQCD can provide an opportunity to evolve the PDF’s to other scales. By convoluting the PDF’s with the fundamental point-like scattering cross sections for partons, one can therefore calculate cross sections for various processes.

In recent years, several groups such as MSTW/MMHT \cite{3,4,5,6},  JR \cite{7}, CTEQ/CT \cite{8,9}, ABM \cite{10,11,12}, and NNPDF \cite{13,14} introduced the PDF's sets by using the HERA data and fixed-target and hadron-collider data. Moreover, in Ref. \cite{15}, the PDF’s have been presented by using a wide variety of Large Hadron Collider (LHC) data and also the combined HERA I + II deep inelastic scattering data set. In an investigation of ultra-high-energy processes, in Refs. \cite{22,23}, the authors have proposed new parameterisations of the proton structure functions by considering the Froissart predictions \cite{21}. The most important benefits of studying these processes are that they confirm HERA investigations and also provide criteria for further investigations of QCD at the Large Hadron Electron Collider (LHeC) in high-energy limit. The kinematic extension of the LHeC is such that it will allow us to check out the nonlinear dynamics at low $x$.
 
HERA ( from 1992 until 2015) combined the neutral current (NC) and charged current (CC) interactions data for $0.045\leq Q^{2}\leq 50000$ $GeV^{2}$ and $6\times 10^{-7}\leq x \leq 0.65$ at values of the inelasticity $0.005\leq y \leq 0.95$ \cite{1}. It operated with an electron beam energy of $E_{e}=27.5GeV$. For most of HERA operations, the proton beam energy was $E_{p} = 920GeV$ and the highest center-of-mass energy in deep inelastic scattering of electrons on protons was $\sqrt{s}= 320 GeV$.  In Ref. \cite{2}, HERA combined the NC and CC differential cross sections, $d\sigma/dQ^{2}$, for $e^{\pm}p$ with predictions from HERAPDF2.0 NLO. Furthermore, HERA  collected $e^{\pm}p$ collision data through the H1 \cite{36} detector, which allowed a measurement of structure functions at $x$ values $6.5\times 10^{-4}\leq x \leq 0.65$ and at $Q^{2}$ values $35\leq Q^{2} \leq 800$ $GeV^{2}$. The differential cross section in terms of the structure function $F_{2}(x,Q^{2})$ and the longitudinal structure function $F_{L}(x,Q^{2})$ at low values of $Q^{2}$ is defined as:
\begin{equation}\label{eq:1}
\frac{d^{2}\sigma}{dxdQ^{2}}=\frac{2\pi \alpha^{2}_{s}Y_{+}}{Q^{4}x}\sigma_{r}(x,Q^{2})=\frac{2\pi \alpha^{2}_{s}Y_{+}}{Q^{4}x}F_{2}(x,Q^{2})\bigg[1-\frac{y^{2}}{Y_{+}}\frac{F_{L}(x,Q^{2})}{F_{2}(x,Q^{2})}\bigg],
\end{equation}
where  $y=Q^{2}/(xs)$ is the inelasticity variable in which $s$  and $Q^{2}$ are the center-of-mass energy squared and the photon virtuality, respectively, and  $Y_{+}=1+(1-y)^{2}$. In the quark-parton model, $F_{2}(x,Q^{2})$ is proportional to the sum of the quark and antiquark distributions. On the other hand, the longitudinal structure function $F_{L}(x,Q^{2})$ is directly sensitive to the gluon density and is nonzero and depends on the strong coupling constant $\alpha_{s}$.

It is known that the pQCD evolution within the DGLAP \cite{16,17,18,19,20,24,25} formalism, at an extremely small $x$, predicts a strong rise of the gluon density and causes a rather singular behavior of the PDF’s which strongly violates the unitary bound (or the Froissart bound \cite{21}).
Consequently, this strong rise increases the proton structure functions in the pQCD. As is well known, the gluon density cannot grow forever, due to this fact that the gluon density constitutes only a limited share of the proton structure functions.
 Based on this bound, the hadronic total cross section cannot grow faster than $\sigma = \frac{\pi}{m_{\pi}} (\ln s)^{ 2}$, where $s$ is the square of the center of mass energy and $m_{\pi}$ is the scale of the strong force.  It is believed that, at high energies, gluon recombination occurs and it can be considered as the mechanism responsible for a possible saturation of gluon densities at small $x$ as well as unitarization of the physical cross sections. By taking into account the recombination processes at small $x$ in a dense system, the growth of the quark and gluon densities can be tamed by screening effects. Accordingly, these effects lead to the appearance of nonlinear terms in the DGLAP evolution equations. Based on  a detailed study at the small $x$ region, Gribov, Levin, Ryskin- Mueller and Qiu (GLR-MQ) \cite{26,27,28}  argued that the physical processes of interaction and recombination of partons are important in the parton cascade at a large value of the parton density. This study leads to the creation of new nonlinear evolution equations, which are the so-called GLR-MQ equations:

\begin{equation}\label{eq:2}
\frac{\partial xq(x,Q^{2})}{\partial \ln Q^{2}}=\frac{\partial xq(x,Q^{2})}{ \partial \ln Q^{2}}\vert_{DGLAP}-\frac{\alpha^{2}_{s}\gamma_{1}}{R^{2}Q^{2}}\left[xg(x,Q^{2})\right]^{2},
\end{equation}
\begin{equation}\label{eq:3}
\frac{\partial xg(x,Q^{2})}{\partial \ln Q^{2}}=\frac{\partial xg(x,Q^{2})}{ \partial \ln Q^{2}}\vert_{DGLAP}-\frac{\alpha^{2}_{s}\gamma_{2}}{R^{2}Q^{2}}\int^{1}_{\chi}\left[yg(y,Q^{2})\right]^{2}\frac{dy}{y},
\end{equation}
where $R$ is the correlation radius between two interacting gluons, $ \gamma_{1}=\frac{27}{160}$ and $\gamma_{2}=\frac{81}{16}$ for $N_{c} =3$, and $\chi = \frac{x}{\rho}$  ($\rho(= 0.01)$ being the boundary condition that the gluon distribution joints smoothly onto the unshadowed region). In these equations, $\frac{\partial xq(x,Q^{2})}{ \partial \ln Q^{2}}\vert_{DGLAP} $ and $\frac{\partial xg(x,Q^{2})}{ \partial \ln Q^{2}}\vert_{DGLAP}$ are obtained from the standard DGLAP evolution equations. The correlation radius $R$ determines the size of the nonlinear terms. Its value depends on how the gluon ladders are coupled to the nucleon or on how the gluons are distributed within the nucleon. On this basis, the value of $R$ is approximately equal to $5GeV^{-1}$ if the gluons are populated across the proton, and is equal to  $2GeV^{-1}$ if the gluons have a hotspot-like structure.

In Ref. \cite{29}, the nonlinear corrections to the distribution functions at low values of $x$ and $Q^{2}$ have been presented by using the parameterisations of $F_{2}(x, Q^{2})$ and consequently  determined the longitudinal structure function $F_{L} (x, Q^{2})$. Indeed, the authors have used a direct method to extract the nonlinear corrections to the ratio of structure functions and to the reduced cross section in the next-to-next-to-leading order (NNLO) approximation. In Ref. \cite{30}, the results of the analytical and numerical analysis of the nonlinear Balitsky– Kovchegov equation has been discussed. One of the important outcomes of this study is the existence of the saturation scale $Q_{s} (x)$ which is a characteristic scale at which the parton recombination effects become important. By considering the nonlinear corrections and using Laplace transform techniques, the authors, in Ref. \cite{31},  have described the determination of the longitudinal structure function $F_{L}(x,Q^{2})$, at the NLO and NNLO approximations, based on  the parameterisation of $F_{2}(x,Q^{2})$ and its derivative with respect to $d\ln Q^{2}$ at low $x$ values. Note that, to perform the calculations, they have used the approximate splitting functions. In Refs. \cite{32,33}, the authors have investigated the phenomenological implications of the parton distribution function sets, with small $x$ resummation, to obtain the longitudinal structure function $F_{L}(x,Q^{2})$ at HERA. Ref. \cite{34} have been devoted to investigate the solutions of the nonlinear evolution equation at the small $x$ region. By using  the Laplace transform technique, the authors, in Ref. \cite{35}, have solved the QCD nonlinear Dokshitzer-Gribov-Lipatov-Altarelli-Parisi (NLDGLAP) and Gribov, Levin, Ryskin- Mueller and Qiu (GLR-MQ) evolution equations at small $x$  and determined the effects of the first nonlinear corrections to the gluon distribution and then obtained the behavior of the gluon distribution. Also, they have shown that the strong rise, corresponding to the linear QCD evolution equations at small $x$ region, can be tamed by screening effects.

In this paper, we intend to investigate the nonlinear corrections to the proton structure functions $F_{2}(x,Q^{2})$ and $F_{L}(x,Q^{2})$ and then to the single differential cross section $d\sigma/dQ^{2}$ and the reduced cross section $\sigma_{r}$. Indeed, we solve the linear DGLAP equations (at the LO and NLO approximations) using the double Laplace transform techniques and then insert the obtained distribution functions in Eqs. (\ref{eq:2}) and (\ref{eq:3}). By doing this, we determine the effects of the nonlinear corrections to the distribution functions. Then, by using AM equations and considering the nonlinear corrections, we obtain the proton structure functions $F_{2}(x,Q^{2})$ and $F_{L}(x,Q^{2})$. Based on these solutions and Eq. (1), we gain the single differential cross section $d\sigma/dQ^{2}$ and the reduced cross section $\sigma_{r}(x,Q^{2})$ at the LO and NLO approximations.

The rest of the present paper is organized as follows:  In section II, at the LO and NLO approximations, we present the solutions to the linear DGLAP evolution equations by applying the double Laplace transform method and then obtain the PDF's. In this section, by considering the GLR-MQ and AM equations, we obtain the nonlinear corrections to the PDF's and the proton structure functions $F_{2}(x,Q^{2})$ and $F_{L}(x,Q^{2})$. In section III, our numerical results of these corrections to the proton structure functions, the single differential cross section $d\sigma/dQ^{2}$ and the reduced cross section $\sigma_{r}(x,Q^{2})$ are presented in interval of $8.5 \leq Q^{2} \leq 5000 GeV^{2}$ and the Bjorken scale $10^{-5} \leq x \leq 1$, and then are compared with the available H1 data \cite{1,2,36,42}, CT18 \cite{15} and the results of Refs. \cite{37,38,29,41}. In section IV, we conclude our presentation. The Appendix includes the kernels and their transformations in the Laplace $s$-space and $u$-space.

\section*{II. METHOD}
The proton structure functions $F_{2}(x,Q^{2})$ and $F_{L}(x,Q^{2})$ are directly related to the singlet and gluon distributions and their behavior can be predicted by AM equations \cite{44}. A formula for the proton structure functions as a convolution integral over the singlet $F_{s}(x,Q^{2})$ and gluon $G(x,Q^{2})$ distribution functions takes the following form:
\begin{equation}\label{eq:4} 
F_{k}(x,Q^{2})=C_{k,ns}(x,Q^{2})\otimes F_{ns}(x,Q^{2})+\langle e^{2}\rangle\bigg[ C_{k,s}(x,Q^{2}) \otimes F_{s}(x,Q^{2})+C_{k,g}(x,Q^{2})\otimes G(x,Q^{2})\bigg],\ \ k=2\ \mbox{and}\ L,
\end{equation}
where $C_{k,j}$'s $(j = ns,s$ and $g)$ are the coefficient functions (given explicitly in the Appendix) and $\langle e^{2}\rangle=\frac{1}{n_{f}}\sum_{k=1}^{n_{f}}e_{q_{k}}^{2}$. The linear coupled DGLAP integral-differential equations are as follows \cite{24,17,18,19,20,22,23}:

\begin{equation}\label{eq:5}
 \frac{\partial F_{s}(x,Q^{2})}{\partial \ln Q^{2}}=\frac{\alpha_{s}(Q^{2})}{2\pi}\bigg[P_{qq}(x,Q^{2})\otimes F_{s}(x,Q^{2})+2n_{f}P_{qg}(x,Q^{2})\otimes G(x,Q^{2})\bigg],\
\end{equation}
\begin{equation}\label{eq:6}
\frac{\partial G(x,Q^{2})}{\partial \ln Q^{2}}=\frac{\alpha_{s}(Q^{2})}{2\pi}\bigg[P_{gq}(x,Q^{2})\otimes F_{s}(x,Q^{2})+P_{gg}(x,Q^{2})\otimes G(x,Q^{2})\bigg], \quad
\end{equation}
\begin{equation}\label{eq:7}
\frac{\partial F_{ns}(x,Q^{2})}{\partial \ln Q^{2}}=\frac{\alpha_{s}(Q^{2})}{2\pi} P_{nsqq}(x,Q^{2})\otimes F_{ns}(x,Q^{2}),\qquad\qquad\qquad\qquad\qquad\
\end{equation}
where $\alpha_{s}$ is the running strong coupling constant, $F_{ns}(x,Q^{2})$ is the nonsinglet distribution function and $P_{ab}(x,Q^{2})$'s  are the Altarelli-Parisi splitting functions which have the following form:
\begin{equation}
P_{ab}(x,Q^{2})=P^{(0)}_{ab}(x)+\frac{\alpha_{s}(Q^{2})}{2\pi}P^{(1)}_{ab}(x)+...\ .
\end{equation}
Within the MS-scheme, the standard representation of the QCD running coupling constant $\alpha_{s}$ in the LO and NLO approximations have the forms:
\begin{equation}
\alpha_{s}^{LO}(t)=\frac{4\pi}{\beta_{0}t},\qquad\qquad\qquad
\end{equation}
\begin{equation}
\alpha_{s}^{NLO}(t)=\frac{4\pi}{\beta_{0}t}\left(1-\frac{\beta_{1}\ln t}{\beta_{0}^{2}t}\right),
\end{equation}
where $\beta_{0}=\left(11-2/3n_{f}\right)$, $\beta_{1}=\left(102-38/3n_{f}\right)$ and $t=\ln(Q^{2}/\Lambda^{2})$ in which $\Lambda$ is
the QCD cut-off parameter. Here, $\Lambda$ is considered $0.192$ and $0.146$ (for $n_{f}=4$ and $n_{f}=5$) and also $0.269$ and $0.184$ (for $n_{f}=4$ and $n_{f}=5$) at the LO and NLO approximations, respectively.

In Eqs. (\ref{eq:4}-\ref{eq:7}), the symbol $\otimes$ represents the convolution integral which is defined as $f(x)\otimes h(x)=\int_{x}^{1}f(y)h(x/y)dy/y$. To solve these equations, we use here the Laplace transform method. For this aim, we insert the variables $x=\exp(-v)$, $y=\exp(-w)$ and $\tau(Q^{2},Q_{0}^{2})=\frac{n}{4\pi}\int_{Q^{2}_{0}}^{ Q^{2}} \alpha_{s}(Q{'}^{2})d\ln(Q{'}^{2})$ into the DGLAP equations (\ref{eq:5}-\ref{eq:7}) as follows:
\begin{equation}\label{eq:8}
 \frac{\partial \hat{F}_{s}(v,\tau)}{\partial\tau}=\frac{2}{n}\bigg[\int_{0}^{v}\hat{P}_{qq}(v-w,\tau) \hat{F}_{s}(w,\tau)dw+\int_{0}^{v}4n_{f}\hat{P}_{qg}(v-w,\tau)\hat{G}(w,\tau)dw\bigg],\
\end{equation}
\begin{equation}\label{eq:9}
\frac{\partial \hat{G}(v,\tau)}{\partial\tau}=\frac{2}{n}\bigg[\int_{0}^{v}\bigg[\hat{P}_{gq}(v-w,\tau) \hat{F}_{s}(w,\tau)dw+\int_{0}^{v}\hat{P}_{gg}(v-w,\tau) \hat{G}(w,\tau)dw\bigg], \quad
\end{equation}
\begin{equation}\label{eq:10}
\frac{\partial \hat{F}_{ns}(v,\tau)}{\partial \tau}= \frac{2}{n}\int_{0}^{v}\hat{P}_{nsqq}(v-w)\hat{F}_{ns}(w,\tau)dw.\qquad\qquad\qquad\qquad\qquad\
\end{equation}
In above equations $\hat{H}(v,\tau)\equiv H(\exp(-v),\tau)$ and $n$ is an integer.

The convolution theorem for Laplace transforms allows us to rewrite the right-hand sides of Eqs. (\ref{eq:8}-\ref{eq:10}) by considering the fact that the Laplace transform of the convolution factors is simply the ordinary product of the Laplace transform of the factors. Using the Laplace transform method, we can turn the convolution equations at the LO and NLO approximations from $v$-space and $\tau$-space into $s$-space and $u$-space, respectively, and then solve them straightforwardly in $s$-space and $u$-space as:
\begin{equation}\label{eq:11}
f^{(i)}(s,u)=k_{ff}^{(i)}(s,u)f^{(i)}(s,0)+k_{fg}^{(i)}(s,u)g^{(i)}(s,0),
\end{equation}
\begin{equation}\label{eq:12}
g^{(i)}(s,u)=k_{gf}^{(i)}(s,u)f^{(i)}(s,0)+k_{gg}(s,u)^{(i)}g^{(i)}(s,0),
\end{equation}
\begin{equation}\label{eq:13}
f_{ns}^{(i)}(s,\tau)=k_{ff,ns}^{(i)}(s,\tau)f_{ns}^{(i)}(s,0),\quad i=\mbox{LO or NLO},
\end{equation}
where $f(s,0)$, $g(s,0)$ and $f_{ns}(s,0)$ are respectively the singlet, gluon and nonsinglet distribution functions at initial scale $\tau=0$ (i.e., $Q^{2}_{0}$). Note that, in Eqs. (\ref{eq:11}-\ref{eq:13}), (i) The kernels $k_{ij}(s,u)$'s at the LO and NLO approximations are given in the Appendix. (ii) $\mathcal{L}[\mathcal{L}[\hat{H}(v,\tau),v,s],\tau,u]= h(s,u)$. (iii) To simplify calculations at the NLO approximation, we use an appropriate approximation for $\alpha_{s}$ where have been utilized in Refs. \cite{45,46,47}. 

The main aim of this paper is to find a solution for the nonlinear DGLAP evolution equations in the saturation region and this can be done by using Eqs. (\ref{eq:2}) and (\ref{eq:3}). The latter equations slow down the $Q^{2}$ evolution of quarks and gluons rather than the standard DGLAP behavior, therefore, by using them, one can respectively write the singlet $F_{s}(x,Q^{2})$ and gluon $G(x,Q^{2})$ distribution functions as follows: 
\begin{equation}\label{eq:14}
\frac{\partial F_{s}(x,Q^{2})}{\partial \ln Q^{2}}=\frac{\partial F_{s}(x,Q^{2})}{ \partial \ln Q^{2}}\vert_{DGLAP}-\frac{2n_{f}\alpha^{2}_{s}\gamma_{1}}{R^{2}Q^{2}}\left[G(x,Q^{2})\right]^{2},
\end{equation}
\begin{equation}\label{eq:15}
\frac{\partial G(x,Q^{2})}{\partial \ln Q^{2}}=\frac{\partial G(x,Q^{2})}{ \partial \ln Q^{2}}\vert_{DGLAP}-\frac{\alpha^{2}_{s}\gamma_{2}}{R^{2}Q^{2}}\int^{1}_{\chi}\left[G(y,Q^{2})\right]^{2}\frac{dy}{y}.
\end{equation}
Now, by using the variable changes $x=\exp(-v)$ and $y=\exp(-w)$ in Eq. (\ref{eq:14}), $x=\exp(-v+\ln(\rho))$ and $y=\exp(-w)$ in Eq. (\ref{eq:15}) and by using again the relation $\tau(Q^{2},Q_{0}^{2})=\frac{n}{4\pi}\int_{Q^{2}_{0}}^{ Q^{2}} \alpha_{s}(Q{'}^{2})d\ln(Q{'}^{2})$, we rewrite respectively Eqs. (\ref{eq:14}) and (\ref{eq:15}) as the following forms; 
\begin{equation}\label{eq:16}
\frac{\partial \hat{F}_{s}(v,z)}{\partial \tau}=\frac{\partial \hat{F}_{s}(v,\tau)}{ \partial z}\vert_{DGLAP}-a_{1}e^{-b_{1}\tau}\left[\hat{G}(v,\tau)\right]^{2},
\end{equation}
\begin{equation}\label{eq:17}
\frac{\partial \hat{G}(v-\ln(\rho),\tau)}{\partial \tau}=\frac{\partial \hat{G}(v-\ln(\rho),\tau)}{ \partial \tau}\vert_{DGLAP}-a_{2}e^{-b_{2}\tau}\int^{v}_{0}\left[\hat{G}(w,\tau)\right]^{2}dw.
\end{equation}
It should be stated that, in the above equations, we have used two suitable approximations $\frac{2n_{f}4\pi\alpha_{s}\gamma_{1}}{nR^{2}Q^{2}}=a_{1}\exp^{-b_{1}\tau}$ and $\frac{4\pi\alpha_{s}\gamma_{2}}{nR^{2}Q^{2}}=a_{2}\exp^{-b_{1}\tau}$ (the errors of these approximations are given in Table (1)). Therefore, we can turn the above equations from $v$-space and $\tau$-space into $s$-space and $u$-space, respectively. In the $v$-space, it has been defined  $\mathcal{L}\left[\int^{v}_{0}\left[\hat{G}(w,z)\right]^{2}dw,v,s\right]= \frac{1}{s}\mathcal{L}\left[\left[\hat{G}(v,\tau)\right]^{2},v,s\right]$ to be less than $\frac{1}{s}\left[\hat{G}(s,\tau)\right]^{2}$ \cite{35}. On this basis, we obtain the Laplace transform of Eqs. (\ref{eq:16}) and (\ref{eq:17}) as follows:
\begin{equation}\label{eq:18}
uf_{s}(s,u)-f_{s}(s,0)=\left(uf_{s}(s,u)-f_{s}(s,0)\right)\vert_{DGLAP}-a_{1}\left[g(s,u+b_{1})\right]^{2},
\end{equation}
\begin{equation}\label{eq:19}
ug(s,u)-g(s,0)=\left(ug(s,u)-g(s,0)\right)\vert_{DGLAP}-\frac{a_{2}}{s\rho^{-s}}\left[g(s,u+b_{2})\right]^{2}.
\end{equation}
Now, by inserting Eqs. (\ref{eq:11}) and (\ref{eq:12}) into Eqs. (\ref{eq:18}) and (\ref{eq:19}), we can write respectively the singlet and gluon distribution functions at the LO and NLO approximations as follows:
\begin{equation}\label{eq:20}
f^{(i)}_{s}(s,u)=k_{ff}^{(i)}(s,u)f_{s0}^{(i)}(s)+k_{fg}^{(i)}(s,u)g_{0}^{(i)}(s)-\frac{a_{1}}{u}\left[g^{(i)}(s,u+b_{i})\right]^{2},
\end{equation}

\begin{equation}\label{eq:21}
g^{(i)}(s,u)=k_{gf}^{(i)}(s,u)f_{s0}^{(i)}(s)+k_{gg}^{(i)}(s,u)g_{0}^{(i)}(s)-\frac{a_{2}}{s\rho^{-s}u}\left[g^{(i)}(s,u+b_{2})\right]^{2},\quad i=\mbox{LO or NLO}.
\end{equation}
To obtain the distribution functions of singlet and gluon in Laplace space, we first solve Eq. (\ref{eq:21}) and then insert its solution into Eq. (\ref{eq:20}). To do this, we use the fact that the value of $a_{2}$ is smaller than one, so we can rewrite this equation in terms of a power series of $a_{2}$. It should be noted that, as much as the value of $n$  is chosen larger than one the series converges faster. Accordingly, one can rewrite the gluon distribution function (Eq. (\ref{eq:21})) as follows:
\begin{equation}\label{eq:22}
g^{(i)}_{1}(s,u)=r_{11}^{(i)} f_{s0}^{(i)}(s)+r_{21}^{(i)} g_{0}^{(i)}(s),
\end{equation}
\begin{equation}\label{eq:23}
g^{(i)}_{2}(s,u)=r_{12}^{(i)} f_{s0}^{(i)}(s)+r_{22}^{(i)}g_{0}^{(i)}(s)+r_{32}^{(i)}f_{s0}^{(i)}(s)g_{0}^{(i)}(s)+r_{42}^{(i)}f_{s0}^{(i)2}(s)+r_{52}^{(i)}g_{0}^{(i)2}(s),
\end{equation}
and
\begin{equation}\label{eq:24}
g^{(i)}_{j}(s,u)=r_{1j}^{(i)} f_{s0}^{(i)}(s)+r_{2j}^{(i)}g_{0}^{(i)}(s)+r_{3j}^{(i)}f_{s0}^{(i)}(s)g_{0}^{(i)}(s)+r_{4j}^{(i)}f_{s0}^{(i)2}(s)+r_{5j}^{(i)}g_{0}^{(i)2}(s)+\mathcal{O}(f_{s0})+\mathcal{O}(g_{0})+\mathcal{O}(f_{s0}g_{0}),
\end{equation}
where indices $1,2 ,...,j$ represent the number of expansion terms and $r_{ij}$'s are the coefficients in terms of $s$ and $u$ (given until $j=2$ in the Appendix). Now, to obtain the singlet distribution function (Eq. (\ref{eq:20})), we insert above equation (\ref{eq:24}) into Eq. (\ref{eq:20}) and then obtain the following forms:
\begin{equation}\label{eq:241}
f^{(i)}_{s1}(s,u)=q_{11}^{(i)} f_{s0}^{(i)}(s)+r_{21}^{(i)}g_{0}^{(i)}(s)+r_{31}^{(i)}f_{s0}^{(i)}(s)g_{0}^{(i)}(s)+q_{41}^{(i)}f_{s0}^{(i)2}(s)+q_{51}^{(i)}g_{0}^{(i)2}(s),
\end{equation}
$$
f^{(i)}_{s2}(s,u)=q_{12}^{(i)} f_{s0}^{(i)}(s)+r_{22}^{(i)}g_{0}^{(i)}(s)+r_{32}^{(i)}f_{s0}^{(i)}(s)g_{0}^{(i)}(s)+q_{42}^{(i)}f_{s0}^{(i)2}(s)+q_{52}^{(i)}g_{0}^{(i)2}(s)+q_{62}^{(i)} f_{s0}^{(i)3}(s)+r_{72}^{(i)}f_{s0}^{(i)2}g_{0}^{(i)}(s)
$$
\begin{equation}\label{eq:242}
+r_{82}^{(i)}f_{s0}^{(i)}(s)g_{0}^{(i)2}(s)+q_{92}^{(i)}g_{0}^{(l)3}(s)+q_{102}^{(i)}f_{s0}^{(i)2}(s)g_{0}^{(i)2}(s)+q_{112}^{(i)}f_{s0}^{(i)4}(s)+q_{122}^{(i)}f_{s0}^{(i)3}(s)g_{0}^{(i)}(s)+q_{132}^{(i)}f_{s0}^{(i)}(s)g_{0}^{(i)3}(s)+q_{142}^{(i)}g_{0}^{(i)4}(s),
\end{equation}
and
\begin{equation}\label{eq:25}
f^{(i)}_{sj}(s,u)=q_{1j}^{(i)} f_{s0}^{(i)}(s)+q_{2j}^{(i)}g_{0}^{(i)}(s)+q_{3j}^{(i)}f_{s0}^{(i)}(s)g_{0}^{(i)}(s)+q_{4j}^{(i)}f_{s0}^{(i)2}(s)+q_{5j}^{(i)}g_{0}^{(i)2}(s)+\mathcal{O}(f_{s0})+\mathcal{O}(g_{0})+\mathcal{O}(f_{s0}g_{0}),
\end{equation}
where $q_{ij}$'s are the coefficients in terms of $s$ and $u$ (given until $j=2$ in the Appendix). 
Now by applying the variable changes $x=\exp(-v)$ and $y=\exp(-w)$ in Eq. (\ref{eq:4}), we can take the Laplace transform of this equation. So, this equation in $s$-space reads as follows:
\begin{equation}\label{eq:26}
 f_{k}^{(i)}(s,\tau)=c_{kns}^{(i)}(s,\tau)f_{ns}^{(i)}(s,\tau)+\langle e^{2}\rangle\left\lbrace c_{ks}^{(i)}(s,\tau) f_{s}^{(i)}(s,\tau)+c_{kg}^{(i)}(s,\tau)g^{(i)}(s,\tau)\right\rbrace,\qquad\qquad\quad \
\end{equation}
where the kernels $c_{ij}$'s are given in the Appendix. To solve Eq. (\ref{eq:26}), we have to return Eqs. (\ref{eq:24}) and (\ref{eq:25}) to the usual space $\tau$. For this purpose, we use the Laplace inverse transform. By inserting the inverse of Eqs. (\ref{eq:24}) and (\ref{eq:25}) into Eq. (\ref{eq:26}) and using Eq. (\ref{eq:13}), one can obtain the nonlinear corrections to the proton structure functions in $s$-space as follows:
$$
 f_{kj}^{(i)}(s,\tau)=w_{kns}^{(i)}(s,\tau)f_{ns0}^{(i)}(s)+\bigg[ w_{k1j}^{(i)}(s,\tau) f_{s0}^{(i)}(s)+w_{k2j}^{(i)}(s,\tau)g_{0}^{(i)}(s)+w_{k3j}^{(i)}(s,\tau)f_{s0}^{(i)}(s)g_{0}^{(i)}(s)
 $$
 \begin{equation}\label{eq:27}
 +w_{k4j}^{(i)}(s,\tau)f_{s0}^{(i)2}(s)+w_{k5j}^{(i)}(s,\tau)g_{0}^{(i)2}(s)+\mathcal{O}(f_{s0})+\mathcal{O}(g_{0})+\mathcal{O}(f_{s0}g_{0})\bigg], \ \  i=\mbox{LO or NLO} \ \ \mbox{and} \ \ k=2,L,
\end{equation}
where the kernels $w_{ij}$'s are
\begin{equation}\label{eq:28}
 w_{kns}^{(i)}(s,\tau)=c_{kns}^{(i)}(s,\tau)k^{(i)}_{ffns}(s,\tau),
\end{equation}
\begin{equation}\label{eq:29}
 w_{khj}^{(i)}(s,\tau)=\langle e^{2}\rangle\bigg(c_{ks}^{(i)}(s,\tau)Q_{hj}^{(i)}(s,\tau)+c_{kg}^{(i)}(s,\tau)R_{hj}^{(i)}(s,\tau)\bigg), \ \ \ h=1,2,3,\ldots ,
\end{equation}
in which $Q_{hj}^{(i)}(s,\tau)=\mathcal{L}^{-1}[q_{hj}^{(i)}(s,u),u,\tau]$ and $R_{hj}^{(i)}(s,\tau)=\mathcal{L}^{-1}[r_{hj}^{(i)}(s,u),u,\tau]$. Now, by applying the inverse Laplace transform for Eq. (\ref{eq:27}), the proton structure functions $F_{2}(x,Q^{2})$ and $F_{L}(x,Q^{2})$ in the usual $x$-space can be written as:
$$
 F_{kj}^{(i)}(x,\tau)=W_{kns}^{(i)}(x,\tau)\otimes F_{ns0}^{(i)}(x)+\bigg[ W_{k1j}^{(i)}(x,\tau) \otimes F_{s0}^{(i)}(x)+W_{k2j}^{(i)}(x,\tau)\otimes G_{0}^{(i)}(s)+W_{k3j}^{(i)}(x,\tau)\otimes H^{(i)}_{1}(x)
 $$
 \begin{equation}\label{eq:30}
 +W_{k4j}^{(i)}(x,\tau)\otimes H_{2}^{(i)}(x)+W_{k5j}^{(i)}(x,\tau)H^{(i)}_{3}(x)+...\bigg], \ \  i=1,2 \ \ \mbox{and} \ \ k=2,L,
\end{equation}
where $W_{ij}(x,\tau)=\mathcal{L}^{-1}[w_{ij}(s,\tau),s,v]\mid_{v=\ln(1/x)}$, $H^{(i)}_{1}(x)= \mathcal{L}^{-1} [f_{s0}^{(i)}(s)g_{0}^{(i)}(s),s,v]\mid_{v=\ln(1/x)}$, $H^{(i)}_{2}(x)= \mathcal{L}^{-1} [f_{s0}^{(i)2}(s),s,v]\mid_{v=\ln(1/x)}$ and $H^{(i)}_{3}(x)= \mathcal{L}^{-1} [g_{0}^{(i)2}(s),s,v]\mid_{v=\ln(1/x)}$.

\section*{III. NUMERICAL RESULTS }
In this section, we present our numerical results of the nonlinear corrections to the proton structure functions $F_{2}(x,Q^{2})$ and $F_{L}(x,Q^{2})$, the reduced cross section $\sigma_{r}(x,Q^{2})$ and the differential cross section $d\sigma/dQ^{2}$ obtained by the DGLAP evolution, AM and GLR-MQ equations. In order to present more detailed discussions on our findings, the numerical results for the proton structure functions $F_{2}(x,Q^{2})$ and $F_{L}(x,Q^{2})$ are compared with H1 collaboration data \cite{36}, the parameterisation model of PM \cite{37,38}, the results of  NNLO-BR \cite{29} and the results of CT18 \cite{15} at the NNLO approximations (which the latter have been obtained using a wide variety of high-precision Large Hadron Collider data). Furthermore, the results of the reduced cross section $\sigma_{r}(x,Q^{2})$ and the differential cross section $d\sigma/dQ^{2}$ are compared with the H1 data \cite{1,42,2}. To extract numerical results, we use the published MSTW2008 \cite{5} initial starting functions $F_{s0}(x)$, $G_{0}(x)$ and $F_{ns0} (x)$ at $Q^{2}_{0}= 1 GeV^{2}$ and also consider $n=500$ and $R=5GeV^{-1}$. It should be noted that, in calculations, the uncertainties are due to the PDF’s at initial scale and the errors in table (1) which are shown as the error bars.

In figure (1), we present the $x$-dependence of the longitudinal structure function $F_{L}(x,Q^{2})$ at $Q^{2} = 8.5, 15, 35, 90, 300$ and $800 GeV^{2}$ and compare it with H1 collaboration data \cite{36}. In this figure, blue solid and blue dashed curves are the numerical results of the longitudinal structure function,  including the nonlinear corrctions, at the NLO and LO approximations, respectively, and black solid and black dashed curves are the numerical results of the longitudinal structure function, regardless of the nonlinear corrections, at NLO and LO approximations, respectively. It is seen that the effects of the nonlinear corrections are more noticeable at $x<0.001$.  Figure (2) shows the nonlinear corrections to the longitudinal structure function $F_{L}(x,Q^{2})$ based on the double Laplace transform method at the LO and NLO approximations. In this figure, the results of the nonlinear corrections at  $8.5<Q^{2}<800$ $GeV^2$ are compared with the H1 collaboration data \cite{36}, the parameterisation models of PM \cite{37,38},  NNLO-BR \cite{29} and also CT18 \cite{15} at the NNLO approximation (the CT18 results have been obtained at a fixed value of the invariant mass $W$ as $W = 230 GeV$). 
\begin{figure}[h]
\includegraphics[width=.7\textwidth]{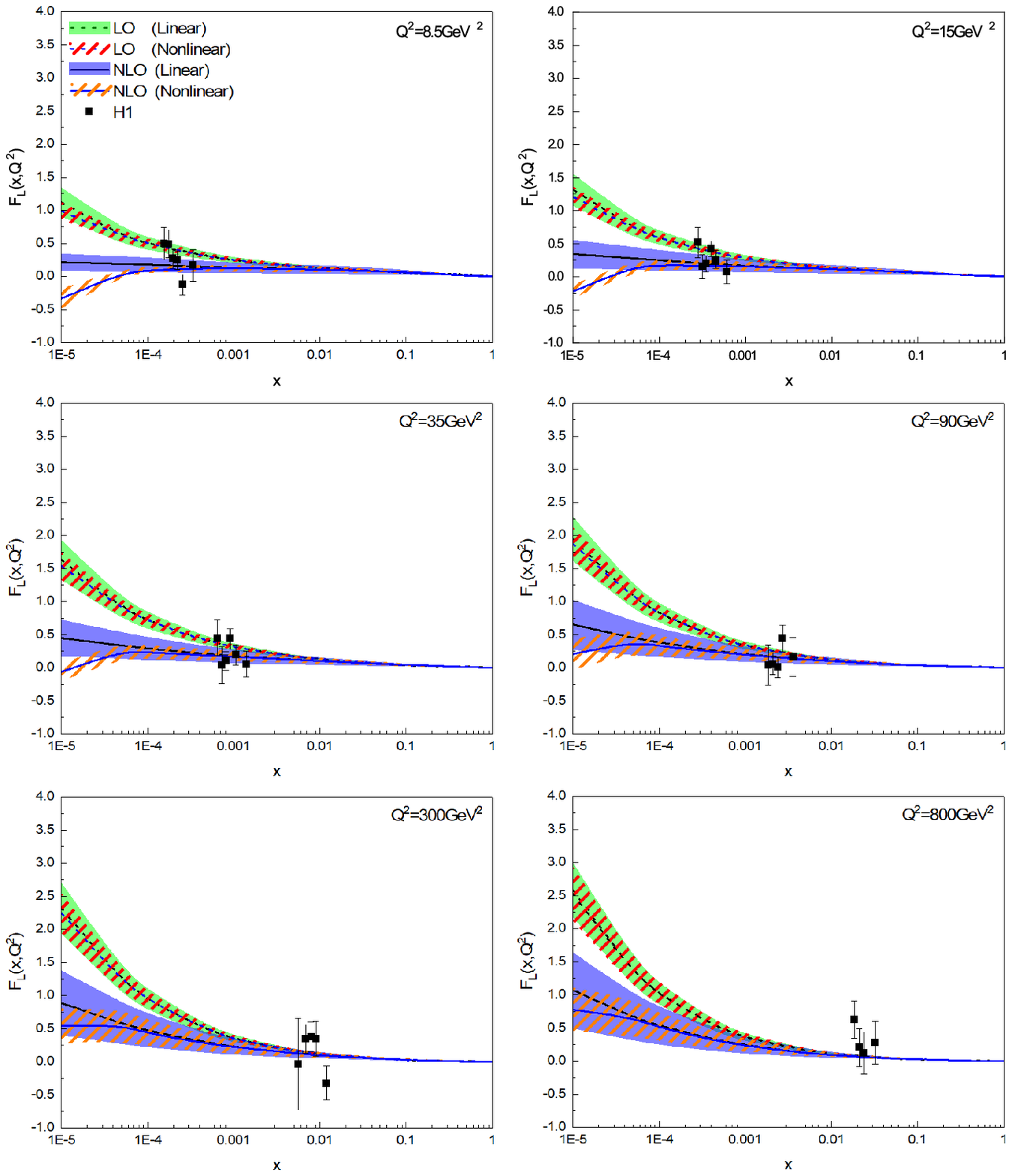}
\begin{center}
\selectfont{\label*{Figure (1): The nonlinear longitudinal structure function $F_{L}(x,Q^{2})$ at the LO (blue dashed curves) and NLO (blue solid curves) approximations in $Q^{2} = 8.5, 15, 35, 90, 300$ and $800$ $GeV^{2}$. The results have been compared with the H1 collaboration data\cite{36}.}}
\end{center}
\end{figure}
\begin{figure}[h]
\begin{center}
\includegraphics[width=.7\textwidth]{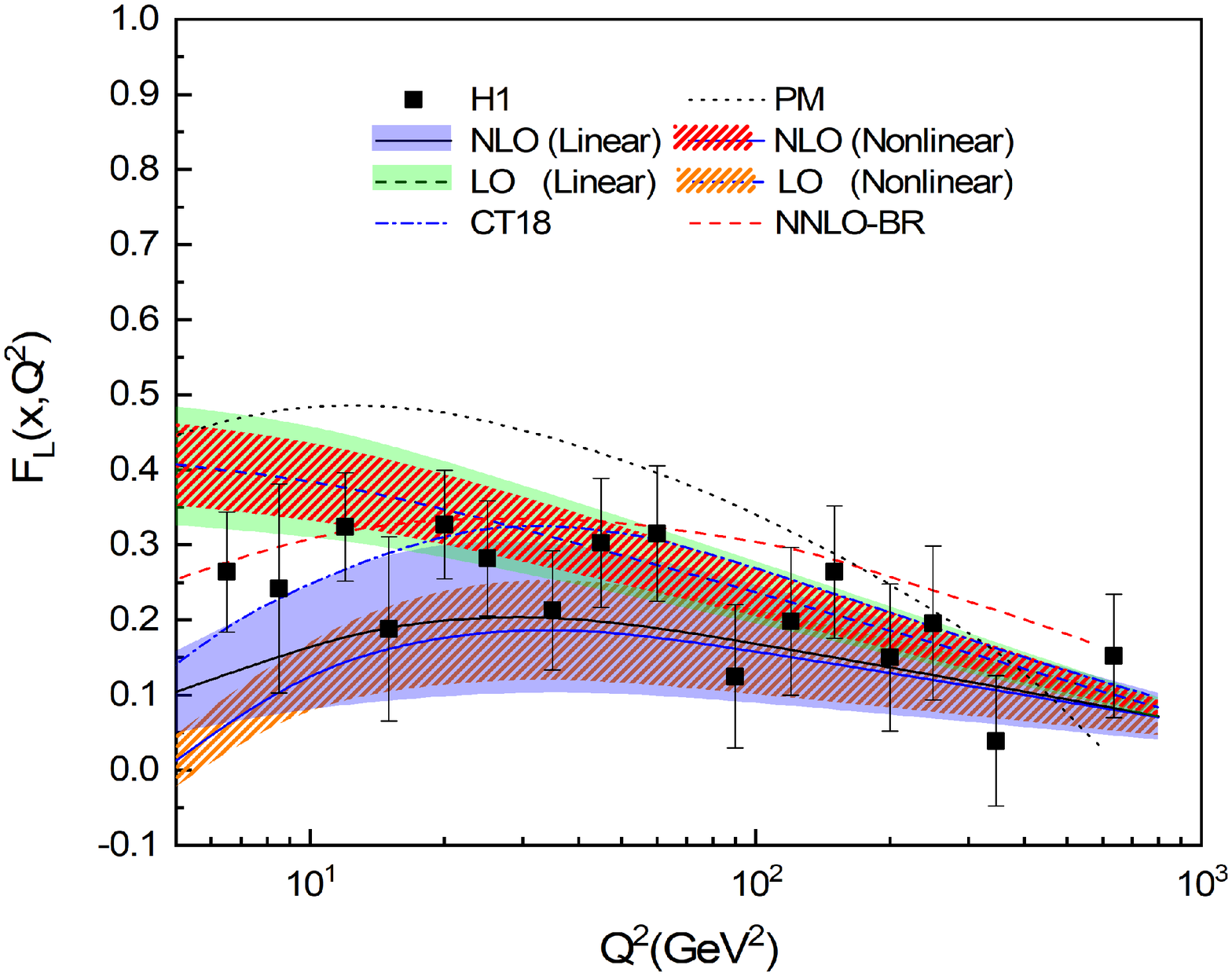}
\end{center}
\begin{center}
\selectfont{\label*{Figure (2): The nonlinear longitudinal structure function $F_{L}(x,Q^{2})$ at the LO and NLO approximations averaged over $x$ at different $Q^{2}$. The nonlinear results at the LO (blue dashed curves) and NLO (blue solid curves) approximations have been compared with the H1 collaboration data \cite{36}, the parameterisation models of PM (dot curves) \cite{37,38}, NNLO-BR \cite{29} (red dashed curves) and CT18  \cite{15} (dashed dot curves) at the NNLO approximation.}}
\end{center}
\end{figure}

In figure (3), the nonlinear corrections to the proton structure function $F_{2}(x,Q^{2})$ at the LO and NLO approximations are presented at $Q^{2} = 8.5, 15, 35, 90, 300, 800 GeV^{2}$. In this figure, our numerical results are compared with the H1 collaboration data \cite{36} and the linear results. Furthermore, in figure (4), the proton structure function $F_{2}(x,Q^{2})$ obtained by using the double Laplace transforms method are compared with the results of Ref. \cite{41} and with the H1 collaboration data \cite{36} at interval $8.5<Q^{2}<800$ $GeV^2$ for different values of $x$. As can be seen in this figure, the effects of nonlinear corrections are noticeable at low $x$ values and also the results are comparable with the experimental data.

\begin{figure}[h]
\begin{center}
\includegraphics[width=.7\textwidth]{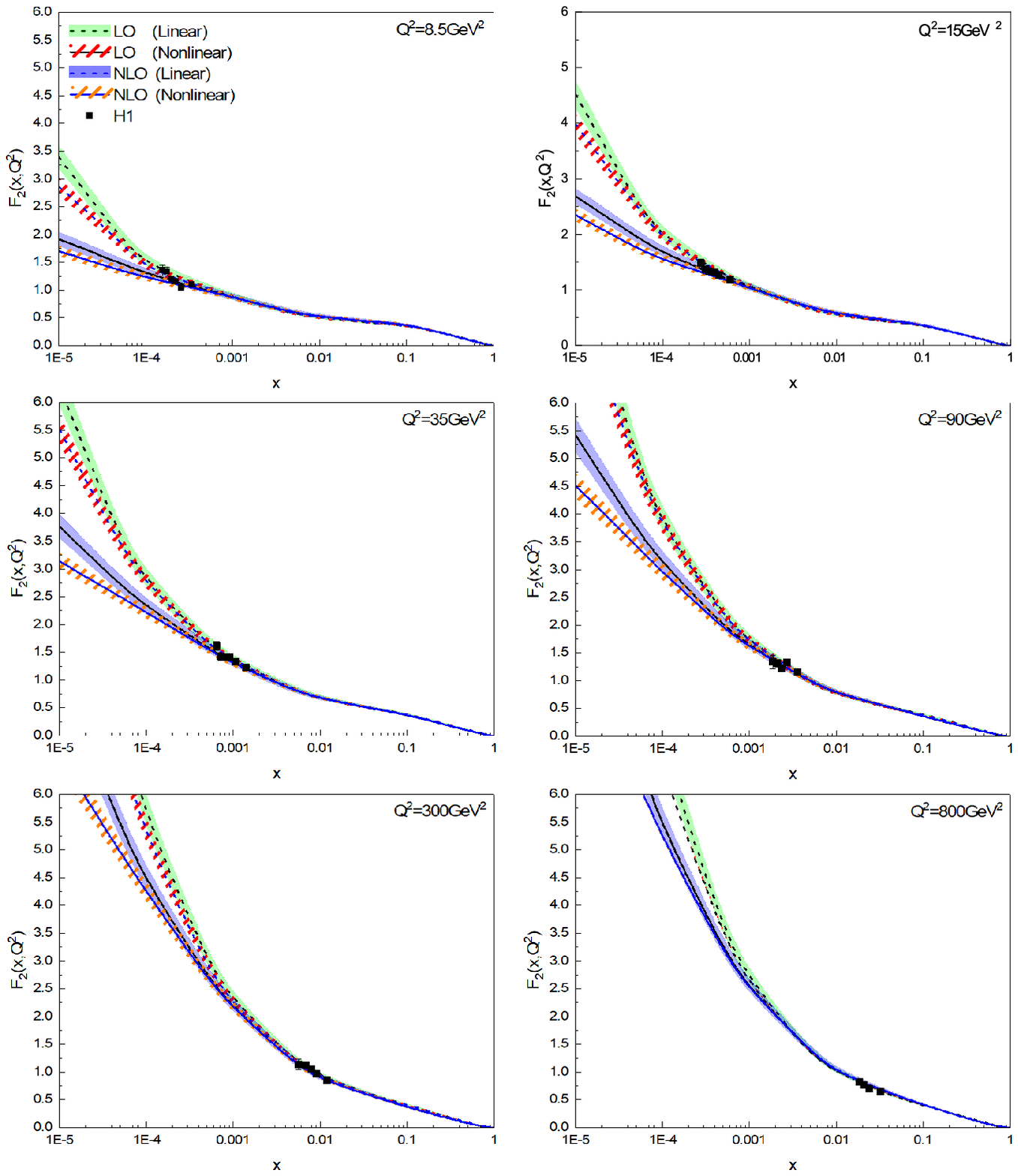}
\end{center}
\begin{center}
\selectfont{\label*{Figure (3): The nonlinear structure function $F_{2}(x,Q^{2})$ at the LO (blue dashed curves) and NLO (blue solid curves) approximations at $Q^{2} = 8.5, 15, 35, 90, 300$ and $800$ $GeV^{2}$. These results have been compared with the H1 collaboration data \cite{36} and the linear results.}}
\end{center}
\end{figure}
\begin{figure}[h]
\begin{center}
\includegraphics[width=.7\textwidth]{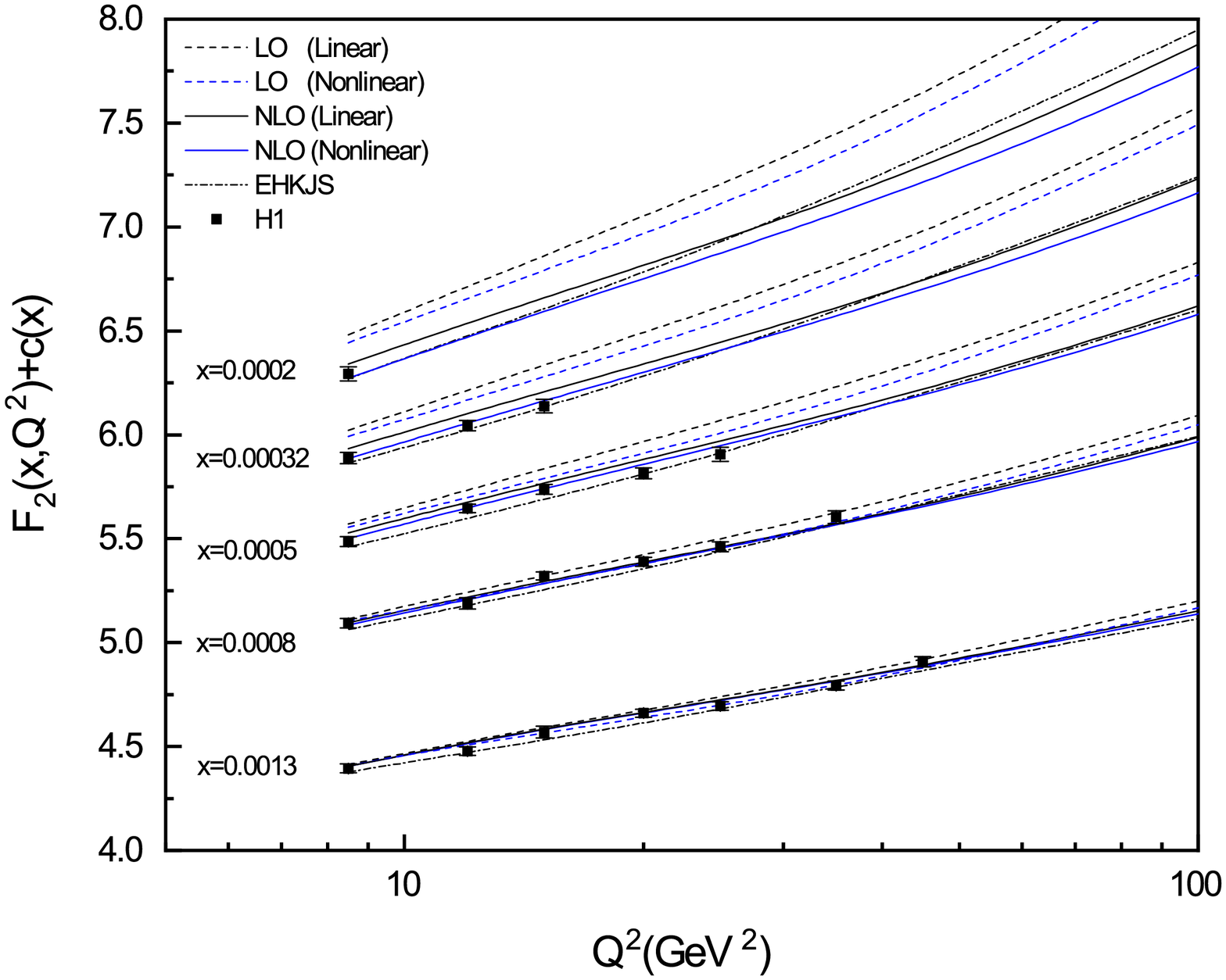}
\end{center}
\begin{center}
\selectfont{\label*{Figure (4): The scale evolution of the proton structure function $F_{2}(x,Q^{2})$ with the nonlinear corrections for fixed values of $x$ (with constants added to separate the curves). The dashed and solid curves respectively show our numerical results at the LO and NLO approximations and the dashed dot curves are the results of EHKJS \cite{41}. The data are from H1 \cite{36} and the error bars are statistical.}}
\end{center}
\end{figure}
\begin{figure}[h]
\begin{center}
\includegraphics[width=.7\textwidth]{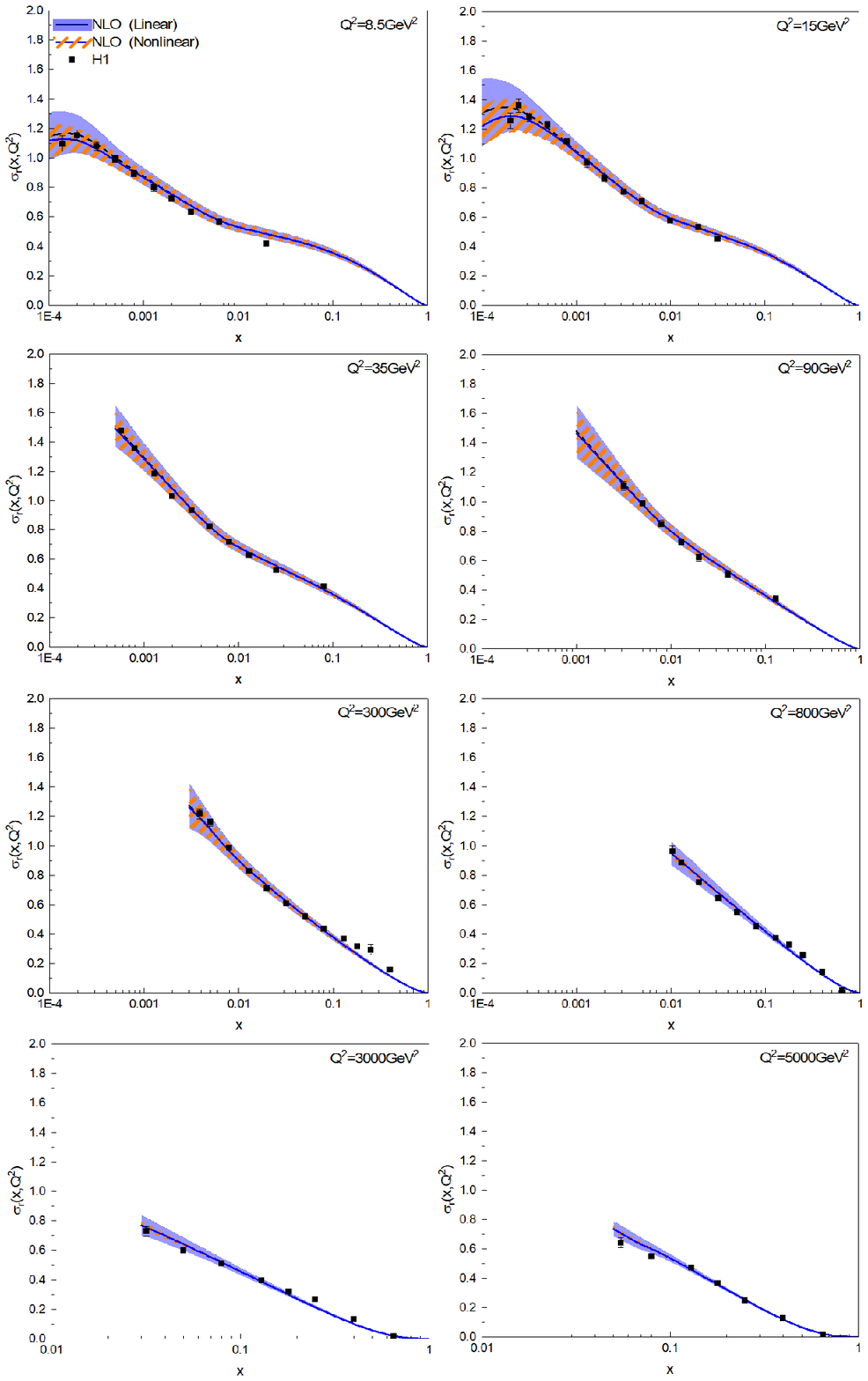}
\end{center}
\begin{center}
\selectfont{\label*{Figure (5): A comparison between the combined HERA I+II NC reduced cross sections $\sigma_{r}(x,Q^{2})$ \cite{1,42} and our numerical results at the NLO approximation with the nonlinear corrections in the scales $Q^{2}=8.5,15, 35, 90, 300,800,3000$ and $5000$ $GeV^{2}$. The behaviour of the corrected cross section at small $x$ is better than that of the uncorrected one.}}
\end{center}
\end{figure}
\begin{figure}[h]
\begin{center}
\includegraphics[width=.8\textwidth]{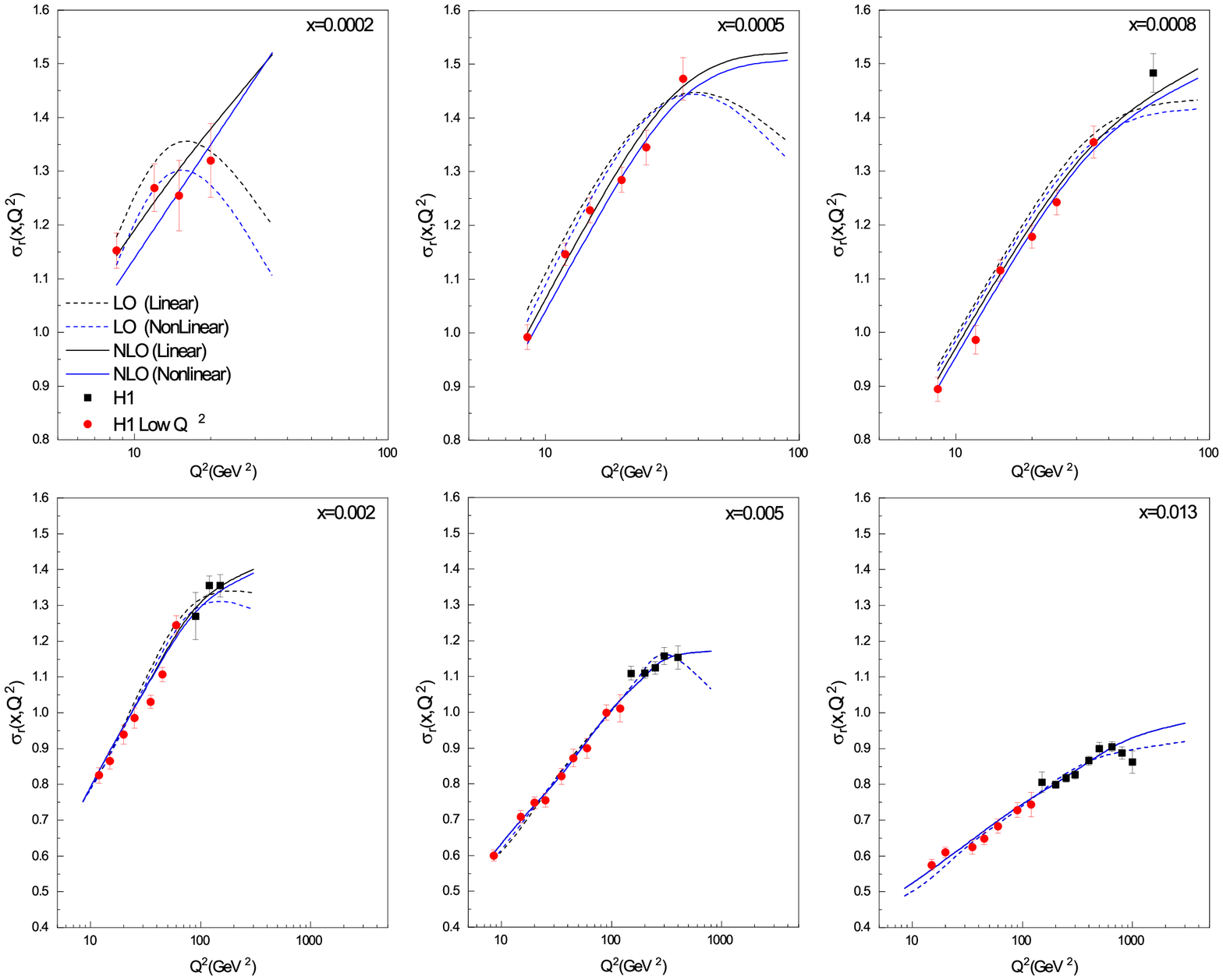}
\end{center}
\begin{center}
\selectfont{\label*{Figure (6): A comparison between the combined HERA I+II  NC reduced cross section $\sigma_{r}(x,Q^{2}) $\cite{2} and our numerical results at the LO and NLO approximations with the nonlinear corrections for various fixed of $x$ as a function of $Q^{2}$. }}
\end{center}
\end{figure} 
Based on Eqs. (1) and (35), we show our numerical results of the NC reduced cross section $\sigma_{r}$ at the NLO approximation in $Q^{2}=8.5,15, 35, 90, 300,800,3000$ and $5000$ $GeV^{2}$ in figure (5). By comparing our results at the NLO approximation with the H1 data, one can conclude that the nonlinear corrections improve the results of the reduced cross section at low to moderate $Q^{2}$ values for low $x$ values and have almost no effect on the results of the reduced cross section at high $x$ values. We note again that the GLR-MQ equations have been obtained from the interaction and recombination of the partons at low $x$ values, as shown in figure (5). In figure (6), a comparison between our numerical results and the combined HERA I+II  NC reduced cross sections $\sigma_{r}(x,Q^{2})$ \cite{2} (with center of mass energy $\sqrt{s} = 319 GeV$ and for $e^{-}p$ at high $Q^{2}$ and low $Q^{2}$ data) is shown for various fixed $x$ as a function of $Q^{2}$ values. 

Figure (7) shows the $Q^{2}$-dependence of the single differential cross section $d\sigma/dQ^{2}$ from the combined HERA I+II NC $e^{-}p$ data \cite{2}. The steep decrease of the differential cross section with increasing $Q^{2}$ is due to the dominating photon exchange cross section which is proportional to $1/Q^{4}$. In this figure, we present the effects of the nonlinear corrections to the differential cross section $d\sigma/dQ^{2}$. Also in this figure, the ratio of the H1 collaboration data \cite{2} to our numerical results is shown in the interval $200\leq Q^{2}\leq 5000$ $GeV^{2}$. At the bottom of this figure, we show the ratio of the nonlinear to linear results of the differential cross section. As can be deduced, the effects of the nonlinear corrections to this differential cross section at low $Q^{2}$ are at least $\%10$, which can be a very important result of the nonlinear corrections.

\begin{figure}[h]
\begin{center}
\includegraphics[width=.7\textwidth]{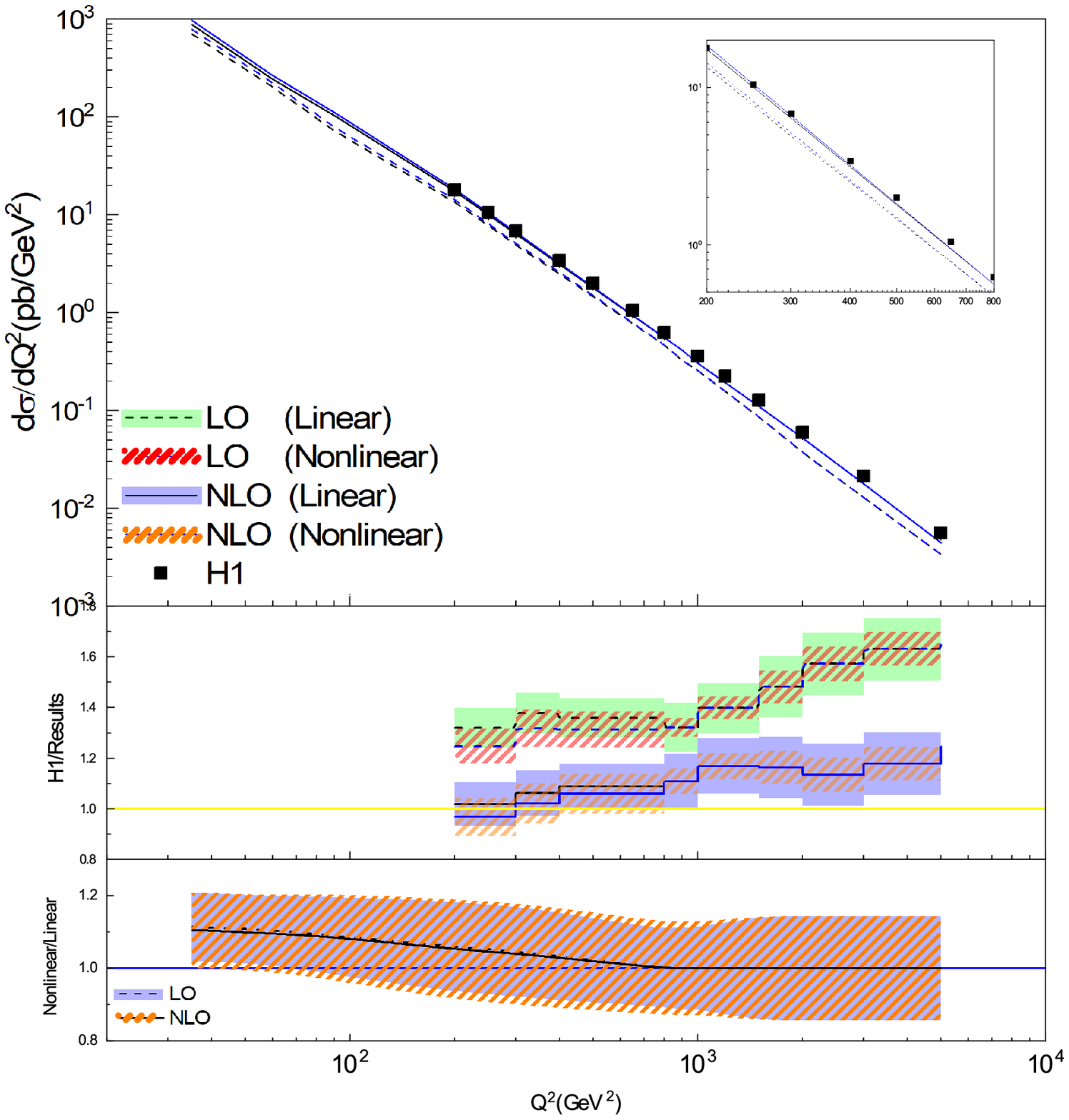}
\end{center}
\begin{center}
\selectfont{\label*{Figure (7): A comparison between the combined HERA I+II  $e^{-} p$ NC differential cross section $d\sigma/dQ^{2}$ \cite{2} and our numerical results with the nonlinear corrections. The ratio of H1 data to the numerical results as well as the results of the nonlinear corrections to the linear results have been displayed. }}
\end{center}
\end{figure}

 \begin{table}[h]
\begingroup
\fontsize{10pt}{12pt}\selectfont{
\label*{ Table (1): The maximum percentage relative errors of exact and approximate values of $\frac{2n_{f}4\pi\alpha_{s}\gamma_{1}}{nR^{2}Q^{2}}=a_{1}\exp^{-b_{1}\tau}$ and $\frac{4\pi\alpha_{s}\gamma_{2}}{nR^{2}Q^{2}}=a_{2}\exp^{-b_{1}\tau}$. }}
\newline
\endgroup
\begingroup
\fontsize{8pt}{12pt}\selectfont{
\begin{tabular} {cccccc}
\toprule[1pt]
&\normal{\head{$M_{c}^{2}\leq Q^{2}<10$ $GeV^{2} $}}&
\normal{\head{$10\leq Q^{2}<M_{b}^{2}$ $GeV^{2} $}}\  & \normal{\head{$M_{b}^{2}\leq Q^{2}<100 $ $GeV^{2}$}}  \ & \normal{\head{$100\leq Q^{2}\leq 500$ $GeV^{2} $}}\ & \normal{\head{$500\leq Q^{2}<5000 $ $GeV^{2} $}}\\ \hline
LO  & $5.6\%$ &  $0.83\%$  &  $4.4\%$& $3.3\%$& $6.5\%$ \\ \\
  & $a_{1}=7.8\times 10^{-4}$ &  $a_{1}=1.4\times 10^{-3}$  &  $a_{1}=2.8\times 10^{-3}$& $a_{1}=9.6\times 10^{-3}$& $a_{1}=5.7\times 10^{-2}$ \\ 
  & $a_{2}=3.0\times 10^{-3}$ &  $a_{2}=5.4\times 10^{-3}$  &  $a_{2}=8.4\times 10^{-3}$& $a_{2}=2.9\times 10^{-2}$& $a_{2}=1.7\times 10^{-1}$ \\
  & $b_{1}=47.69$ &  $b_{1}=57.74$  &  $b_{1}=65.37$& $b_{1}=77.71$& $b_{1}=91.83$ \\ \\

  NLO  & $5.6\%$ &  $0.79\%$  &  $3.8\%$& $3.1\%$& $6.1\%$ \\ \\
  & $a_{1}=7.3\times 10^{-4}$ &  $a_{1}=1.3\times 10^{-3}$  &  $a_{1}=2.3\times 10^{-3}$& $a_{1}=6.3\times 10^{-3}$& $a_{1}=3.6\times 10^{-2}$ \\ 
  & $a_{2}=2.7\times 10^{-3}$ &  $a_{2}=4.9\times 10^{-3}$  &  $a_{2}=6.8\times 10^{-3}$& $a_{2}=2.1\times 10^{-2}$& $a_{2}=1.1\times 10^{-1}$ \\
  & $b_{1}=55.37$ &  $b_{1}=66.66$  &  $b_{1}=74.58$& $b_{1}=87.73$& $b_{1}=102.7$ \\

\bottomrule[1pt]
\end{tabular} }
\endgroup
\end{table}

\hspace*{1em}
\section*{IV. SUMMARY AND CONCLUSION}
We have presented the effects of the nonlinear corrections to the proton structure functions $F_{2}(x,Q^{2})$, $F_{L}(x,Q^{2})$, the single differential cross section $d\sigma/dQ^{2}$ and the NC reduced cross section $\sigma_{r}(x,Q^{2})$ by using Dokshitzer-Gribov-Lipatov-Altarelli-Parisi evolution equations, Altarelli-Martinelli equation, and Gribov-Levin- Ryskin and Mueller equations at the LO and NLO approximations. Indeed, there are various methods to solve the GLR-MQ and AM equations, but in this paper, we have shown that adopting double Laplace transform method is suitable and alternative scheme to solve that equations. The obtained equations are general and require only a knowledge of  the parton distribution functions $F_{s}(x)$, $G(x)$ and $F_{ns}(x)$ at the starting value $Q_{0}^{2}$. Our numerical results have showed that the transition of the proton structure functions from the linear to the nonlinear behavior is considerable and can control the incremental trend of them at low $x$ values. The numerical results of the proton structure functions have been compared with the H1 collaboration data, the results from the CT18 and the NNLO-BR parametrization models at the NLO and NNLO approximations and the results of EHKJS. 
Then we have studied the effects of adding the nonlinear corrections to the reduced cross section at the LO and NLO approximations at various $Q^{2}$ values. By comparing the results obtained for the reduced cross section with the HERA combined data, it can be concluded that these corrections improve the results at low to moderate $Q^{2}$ values for low $x$ values. Furthermore, the numerical results of the single differential cross section have been compared with the H1 collaboration data in a wide range of $Q^{2}$ values. This comparison has showed that the nonlinear corrections have a significant effect on the differential cross section and enhance it at low and moderate $Q^{2}$ values and also demonstrated that these corrections have almost no effect on the differential cross section at high $Q^{2}$. All of the figures clearly show that the extraction procedure provides correct behaviors of the extracted proton structure functions at the LO and NLO approximations. Finally, it should be noted that although the NLO corrections are very small, they often allow to reduce the uncertainties of the predicted cross sections, as one can see by comparing the bands in almost all of the plots presented in the figures.

\section*{Acknowledgements} 
We would like to thank the referee for his/her suggestions  that helped improving the manuscript.

\newpage

\begin{appendix}
\setcounter{equation}{0}
\section{ THE COEFFICIENT FUNCTIONS}
The coefficients $C_{i,j}$ ($i=2,L$ and $j=s,g$) and $C_{ffns}$ in Eq. (\ref{eq:4}) at the LO approximation are as follows \cite{144}:
\begin{eqnarray}
 C_{L,ns}^{(1)}(x,Q^{2})&=&\frac{\alpha_{s}^{(1)}(Q^{2})}{\pi}C_{F}x^{2},\qquad \qquad C^{(1)}_{2,ns}(x,Q^{2}))=\delta(1-x)x,\nonumber\\
C_{L,s}^{(1)}(x,Q^{2})&=&\frac{\alpha_{s}^{(1)}(Q^{2})}{\pi}C_{F}x^{2},\qquad \qquad C^{(1)}_{2,s}(x,Q^{2})=\delta(1-x)x, \nonumber\\
C_{L,g}^{(1)}(x,Q^{2})&=&\frac{\alpha_{s}^{(1)}(Q^{2})}{\pi}2n_{f}x^{2}(1-x),\qquad C^{(1)}_{2,g}(x,Q^{2})=0,
\end{eqnarray}
and at the NLO approximation are as follows:
$$
C_{L,ns}^{(2)}(x,Q^{2})=  \frac{\alpha_{s}^{(2)}(Q^{2})}{\pi}C_{F}x^{2}+\bigg(\frac{\alpha_{s}^{(2)}(Q^{2})}{4\pi}\bigg)^{2}x \bigg[\frac{128}{9}x\times\ln(1-x)^{2}-46.50x\ln(1-x)-84.094\ln(x)\ln(1-x)-37.338
$$
\begin{equation}
+89.53x+33.82x^{2}+x\ln(x)(32.90+18.41\ln(x))-\frac{128}{9}\ln(x)-0.012\delta(x_{1})+\frac{16}{27}n_{f}(6x\ln(1-x)-12x\ln(x)-25x+6)\bigg],
\end{equation}
 
$$
C_{L,s}^{(2)}(x,Q^{2})=C_{L,ns}^{(2)}(x,Q^{2})+\bigg(\frac{\alpha_{s}^{(2)}(Q^{2})}{4\pi}\bigg)^{2}xn_{f}\bigg[(15.94-5.212x)(1-x)^{2}\ln(1-x)+(0.421+1.520x)\ln(x)^{2}
$$
\begin{equation}
+28.09(1-x)\ln(x)-(2.370x^{-1}-19.27)(1-x)^{3}\bigg],
 \end{equation}

$$
C_{L,g}^{(2)}(x,\tau)= \frac{\alpha_{s}^{(2)}(Q^{2})}{\pi}2n_{f}x^{2}(1-x) + \bigg(\frac{\alpha_{s}^{(2)}(Q^{2})}{4\pi}\bigg)^{2}xn_{f}\bigg[(94.74-49.20x)(1-x)\ln(1-x)^{2}+864.8(1-x)\ln(1-x)
$$
\begin{equation}
+1161x\ln(1-x)\ln(x)+60.06(1-x)\ln(x)^{2}+39.66(1-x)\ln(x)-5.333(x^{-1}-1)\bigg].
\end{equation}

$$
C^{(2)}_{2,ns}(x,\tau(Q^{2},Q_{0}^{2}))=\delta(1-x)x+xC_{F}\frac{\tau}{4\pi}\bigg(4D_{1}-3D_{0}-(9+4\zeta_{2})\delta(1-x)-2(1+x)(\ln(1-x)-\ln(x))
$$
\begin{equation}
-4(1-x)^{-1}\ln(x)+6+4x\bigg),
\end{equation}
\begin{equation}
C^{(2)}_{2,s}(x,\tau)=C^{(2)}_{2,ns}(x,\tau),
\end{equation}
\begin{equation}
C^{(2)}_{2,g}(x,\tau)=n_{f}x\frac{\tau}{4\pi}\left((2-4x(1-x))(\ln(1-x)-\ln(x))-2+16x(1-x)\right),
\end{equation}
where $n_{f}$ denotes the number of active massless flavors, $C_{F}=\frac{4}{3}$, $D_{1}=\left[(1-x)^{-1}\ln(1-x)\right]_{+}$ and $D_{0}=\left[(1-x)^{-1}\right]_{+}$.

The coefficients  $k_{i,j}(s,u) $ ( $i,j=f,g$) in Eqs. (\ref{eq:11}-\ref{eq:13}) at the LO approximation are as follows: 
%%%%%%%%%%%%%%%%%%%%%%%%%%%%%%%%%%%%%%%%%%%%%%%%%%%%%%%%%%%%%%%%%%%%%%%%%%%%%%%%%%%%%%%%%%
 \begin{widetext}
 \begin{equation}
 k_{ff}^{(1)}(s,u)=\frac{4 n (n u-\Phi_{g}^{(1)})}{n^2 \left(-\left(T^2-4 u^2\right)\right)+2 \Phi_{f}^{(1)} (\Phi_{g}^{(1)}-2 n u)-4 n \Phi_{g}^{(1)} u+\Phi_{f}^{(1)2}+\Phi_{g}^{(1)^2}},
\end{equation}
\begin{equation}
  k_{fg}^{(1)}(s,u)=-\frac{4 \Theta_{f}^{(1)}}{n \left(T^2-\frac{(-2 n u+\Phi_{f}^{(1)}+\Phi_{g}^{(1)^2})}{n^2}\right)},
\end{equation}
\begin{equation}
   k_{gf}^{(1)}(s,u)=\frac{4 n (n u-\Phi_{f}^{(1)})}{n^2 \left(-\left(T^2-4 u^2\right)\right)+2 \Phi_{f}^{(1)} (\Phi_{g}^{(1)}-2 n u)-4 n \Phi_{g}^{(1)} u+\Phi_{f}^{(1)^2}+\Phi_{g}^{(1)^2}},
\end{equation}
 \begin{equation} 
  k_{gg}^{(1)}(s,u) =-\frac{4 \Theta_{g}^{(1)}}{T^2-\frac{(-2 n u+\Phi_{f}^{(1)}+\Phi_{g}^{(1)})^2}{n^2}},
   \end{equation} 
   \begin{equation} 
  k_{ffns}^{(1)}(s,\tau) = \exp \left(\frac{\tau\Phi_{nsf}^{(1)}}{n}\right),
   \end{equation} 
\end{widetext}
and at the NLO approximation are as follows:
$$
k_{ff}^{(2)}(s,u)=\Bigg[\frac{2 b^3 T n^4}{\Phi_{f}^{(1)}+\Phi_{g}^{(1)}-n (T+2 u)}+\frac{2 b^3 T n^4}{\Phi_{f}^{(1)}+\Phi_{g}^{(1)}+n (T-2 u)}+\frac{2 a \Phi_{f}^{(2)} T^3 n^3}{-2 b n+T n-2 u n+\Phi_{f}^{(1)}+\Phi_{g}^{(1)}}
$$
$$
-\frac{2 a b^2 \Phi_{f}^{(2)} T n^3}{-2 b n+T n-2 u n+\Phi_{f}^{(1)}+\Phi_{g}^{(1)}}-\frac{2 a \Phi_{f}^{(2)} T^3 n^3}{\Phi_{f}^{(1)}+\Phi_{g}^{(1)}-n (T+2 u)}-\frac{2 b^3 \Phi_{f}^{(1)} n^3}{\Phi_{f}^{(1)}+\Phi_{g}^{(1)}-n (T+2 u)}+\frac{2 b^3 \Phi_{g}^{(1)} n^3}{\Phi_{f}^{(1)}+\Phi_{g}^{(1)}-n (T+2 u)}
$$
$$
+\frac{2 a b^2 \Phi_{f}^{(2)} T n^3}{\Phi_{f}^{(1)}+\Phi_{g}^{(1)}-n (T+2 u)}-\frac{2 a \Phi_{f}^{(2)} T^3 n^3}{\Phi_{f}^{(1)}+\Phi_{g}^{(1)}+n (T-2 u)}+\frac{2 b^3 \Phi_{f}^{(1)} n^3}{\Phi_{f}^{(1)}+\Phi_{g}^{(1)}+n (T-2 u)}-\frac{2 b^3 \Phi_{g}^{(1)} n^3}{\Phi_{f}^{(1)}+\Phi_{g}^{(1)}+n (T-2 u)}
$$
$$
+\frac{2 a b^2 \Phi_{f}^{(2)} T n^3}{\Phi_{f}^{(1)}+\Phi_{g}^{(1)}+n (T-2 u)}+\frac{a b^2 \Phi_{f}^{(2)} T n^2}{u-\frac{-2 b n-T n+\Phi_{f}^{(1)}+\Phi_{g}^{(1)}}{2 n}}+\frac{2 a T^2 (\Phi_{f}^{(1)} \Phi_{f}^{(2)}-\Phi_{g}^{(1)} \Phi_{f}^{(2)}+\Theta_{f}^{(2)} \Theta_{g}^{(1)}+\Theta_{f}^{(1)} \Theta_{g}^{(2)}) n^2}{-2 b n+T n-2 u n+\Phi_{f}^{(1)}+\Phi_{g}^{(1)}}
$$
$$
-\frac{a \Phi_{f}^{(2)} T^3 n^2}{u-\frac{-2 b n-T n+\Phi_{f}^{(1)}+\Phi_{g}^{(1)}}{2 n}}-\frac{2 a b^2 \Phi_{f}^{(1)} \Phi_{f}^{(2)} n^2}{-2 b n+T n-2 u n+\Phi_{f}^{(1)}+\Phi_{g}^{(1)}}+\frac{2 a b^2 \Phi_{f}^{(2)} \Phi_{g}^{(1)} n^2}{-2 b n+T n-2 u n+\Phi_{f}^{(1)}+\Phi_{g}^{(1)}}
$$
$$
-\frac{4 a b^2 \Theta_{f}^{(2)} \Theta_{g}^{(1)} n^2}{-2 b n+T n-2 u n+\Phi_{f}^{(1)}+\Phi_{g}^{(1)}}-\frac{2 a b^2 \Phi_{f}^{(1)} \Phi_{f}^{(2)} n^2}{\Phi_{f}^{(1)}+\Phi_{g}^{(1)}-n (T+2 u)}+\frac{2 a b^2 \Phi_{f}^{(2)} \Phi_{g}^{(1)} n^2}{\Phi_{f}^{(1)}+\Phi_{g}^{(1)}-n (T+2 u)}-\frac{4 a b^2 \Theta_{f}^{(1)} \Theta_{g}^{(2)} n^2}{\Phi_{f}^{(1)}+\Phi_{g}^{(1)}-n (T+2 u)}
$$
$$
+\frac{2 a T^2 (\Phi_{f}^{(1)} \Phi_{f}^{(2)}-\Phi_{g}^{(1)} \Phi_{f}^{(2)}+\Theta_{f}^{(2)} \Theta_{g}^{(1)}+\Theta_{f}^{(1)} \Theta_{g}^{(2)}) n^2}{\Phi_{f}^{(1)}+\Phi_{g}^{(1)}-n (T+2 u)}+\frac{2 a b^2 \Phi_{f}^{(1)} \Phi_{f}^{(2)} n^2}{\Phi_{f}^{(1)}+\Phi_{g}^{(1)}+n (T-2 u)}-\frac{2 a b^2 \Phi_{f}^{(2)} \Phi_{g}^{(1)} n^2}{\Phi_{f}^{(1)}+\Phi_{g}^{(1)}+n (T-2 u)}
$$
$$
+\frac{4 a b^2 \Theta_{f}^{(1)} \Theta_{g}^{(2)} n^2}{\Phi_{f}^{(1)}+\Phi_{g}^{(1)}+n (T-2 u)}-\frac{2 a T^2 (\Phi_{f}^{(1)} \Phi_{f}^{(2)}-\Phi_{g}^{(1)} \Phi_{f}^{(2)}+\Theta_{f}^{(2)} \Theta_{g}^{(1)}+\Theta_{f}^{(1)} \Theta_{g}^{(2)}) n^2}{\Phi_{f}^{(1)}+\Phi_{g}^{(1)}+n (T-2 u)}+\frac{a b^2 \Phi_{f}^{(2)} \Phi_{g}^{(1)} n}{u-\frac{-2 b n-T n+\Phi_{f}^{(1)}+\Phi_{g}^{(1)}}{2 n}}
$$
$$
+\frac{a T^2 (\Phi_{f}^{(1)} \Phi_{f}^{(2)}-\Phi_{g}^{(1)} \Phi_{f}^{(2)}+\Theta_{f}^{(2)} \Theta_{g}^{(1)}+\Theta_{f}^{(1)} \Theta_{g}^{(2)}) n}{u-\frac{-2 b n-T n+\Phi_{f}^{(1)}+\Phi_{g}^{(1)}}{2 n}}-\frac{a b^2 \Phi_{f}^{(1)} \Phi_{f}^{(2)} n}{u-\frac{-2 b n-T n+\Phi_{f}^{(1)}+\Phi_{g}^{(1)}}{2 n}}-\frac{2 a b^2 \Theta_{f}^{(2)} \Theta_{g}^{(1)} n}{u-\frac{-2 b n-T n+\Phi_{f}^{(1)}+\Phi_{g}^{(1)}}{2 n}}
$$
$$
-\frac{1}{-2 b n+T n-2 u n+\Phi_{f}^{(1)}+\Phi_{g}^{(1)}}\bigg[2 a b (2 \Phi_{f}^{(2)} \Theta_{f}^{(1)} \Theta_{g}^{(1)}-2 \Phi_{g}^{(2)} \Theta_{f}^{(1)} \Theta_{g}^{(1)}-\Phi_{f}^{(1)} \Theta_{f}^{(2)} \Theta_{g}^{(1)}+\Phi_{g}^{(1)} \Theta_{f}^{(2)} \Theta_{g}^{(1)}+n T \Theta_{f}^{(2)} \Theta_{g}^{(1)}
$$
$$
-\Phi_{f}^{(1)} \Theta_{f}^{(1)} \Theta_{g}^{(2)}+\Phi_{g}^{(1)} \Theta_{f}^{(1)} \Theta_{g}^{(2)}-n T \Theta_{f}^{(1)} \Theta_{g}^{(2)}) n\bigg]
-\frac{1}{-2 b n+T n-2 u n+\Phi_{f}^{(1)}+\Phi_{g}^{(1)}}\bigg[2 a T (2 \Phi_{f}^{(2)} \Theta_{f}^{(1)} \Theta_{g}^{(1)}-2 \Phi_{g}^{(2)} \Theta_{f}^{(1)} \Theta_{g}^{(1)}
$$
$$
-(\Phi_{f}^{(1)}-\Phi_{g}^{(1)}) (\Theta_{f}^{(2)} \Theta_{g}^{(1)}+\Theta_{f}^{(1)} \Theta_{g}^{(2)})) n\bigg]
+\frac{2 a T (2 \Phi_{f}^{(2)} \Theta_{f}^{(1)} \Theta_{g}^{(1)}-2 \Phi_{g}^{(2)} \Theta_{f}^{(1)} \Theta_{g}^{(1)}-(\Phi_{f}^{(1)}-\Phi_{g}^{(1)}) (\Theta_{f}^{(2)} \Theta_{g}^{(1)}+\Theta_{f}^{(1)} \Theta_{g}^{(2)})) n}{\Phi_{f}^{(1)}+\Phi_{g}^{(1)}-n (T+2 u)}
$$
$$
-\frac{1}{\Phi_{f}^{(1)}+\Phi_{g}^{(1)}-n (T+2 u)}\bigg[2 b \bigg(n^3 T^3+n^2 (\Phi_{g}^{(1)}-\Phi_{f}^{(1)}) T^2+a n (\Theta_{f}^{(1)} \Theta_{g}^{(2)}-\Theta_{f}^{(2)} \Theta_{g}^{(1)}) T+a (-2 \Phi_{f}^{(2)} \Theta_{f}^{(1)} \Theta_{g}^{(1)}
$$
$$
+2 \Phi_{g}^{(2)} \Theta_{f}^{(1)} \Theta_{g}^{(1)}+(\Phi_{f}^{(1)}-\Phi_{g}^{(1)}) (\Theta_{f}^{(2)} \Theta_{g}^{(1)}+\Theta_{f}^{(1)} \Theta_{g}^{(2)}))\bigg) n\bigg]
+\frac{1}{\Phi_{f}^{(1)}+\Phi_{g}^{(1)}+n (T-2 u)}\bigg[2 a T (2 \Phi_{f}^{(2)} \Theta_{f}^{(1)} \Theta_{g}^{(1)}-2 \Phi_{g}^{(2)} \Theta_{f}^{(1)} \Theta_{g}^{(1)}
$$
$$
-(\Phi_{f}^{(1)}-\Phi_{g}^{(1)}) (\Theta_{f}^{(2)} \Theta_{g}^{(1)}+\Theta_{f}^{(1)} \Theta_{g}^{(2)})) n\bigg]
-\frac{1}{\Phi_{f}^{(1)}+\Phi_{g}^{(1)}+n (T-2 u)}\bigg[2 b \bigg(n^3 T^3+n^2 (\Phi_{f}^{(1)}-\Phi_{g}^{(1)}) T^2+a n (\Theta_{f}^{(1)} \Theta_{g}^{(2)}
$$
$$
-\Theta_{f}^{(2)} \Theta_{g}^{(1)}) T+a (2 \Phi_{f}^{(2)} \Theta_{f}^{(1)} \Theta_{g}^{(1)}-2 \Phi_{g}^{(2)} \Theta_{f}^{(1)} \Theta_{g}^{(1)}-(\Phi_{f}^{(1)}-\Phi_{g}^{(1)}) (\Theta_{f}^{(2)} \Theta_{g}^{(1)}+\Theta_{f}^{(1)} \Theta_{g}^{(2)}))\bigg) n\bigg]
$$
$$
+\frac{1}{u-\frac{-2 b n-T n+\Phi_{f}^{(1)}+\Phi_{g}^{(1)}}{2 n}}\bigg[a b (-2 \Phi_{f}^{(2)} \Theta_{f}^{(1)} \Theta_{g}^{(1)}+2 \Phi_{g}^{(2)} \Theta_{f}^{(1)} \Theta_{g}^{(1)}+\Phi_{f}^{(1)} \Theta_{f}^{(2)} \Theta_{g}^{(1)}-\Phi_{g}^{(1)} \Theta_{f}^{(2)} \Theta_{g}^{(1)}+n T \Theta_{f}^{(2)} \Theta_{g}^{(1)}+\Phi_{f}^{(1)} \Theta_{f}^{(1)} \Theta_{g}^{(2)}
$$
$$
-\Phi_{g}^{(1)} \Theta_{f}^{(1)} \Theta_{g}^{(2)}-n T \Theta_{f}^{(1)} \Theta_{g}^{(2)})\bigg]
+\frac{a T (2 \Phi_{f}^{(2)} \Theta_{f}^{(1)} \Theta_{g}^{(1)}-2 \Phi_{g}^{(2)} \Theta_{f}^{(1)} \Theta_{g}^{(1)}-(\Phi_{f}^{(1)}-\Phi_{g}^{(1)}) (\Theta_{f}^{(2)} \Theta_{g}^{(1)}+\Theta_{f}^{(1)} \Theta_{g}^{(2)}))}{u-\frac{-2 b n-T n+\Phi_{f}^{(1)}+\Phi_{g}^{(1)}}{2 n}}\Bigg]
$$
\begin{equation}
\Bigg/\bigg[2 b n^3 T \left(T^2-b^2\right)\bigg],
\end{equation}
$$
k_{fg}^{(2)}(s,u)=\Bigg[\frac{4 b^3 \Theta_{f}^{(1)} n^3}{\Phi_{f}^{(1)}+\Phi_{g}^{(1)}-n (T+2 u)}-\frac{4 b T^2 \Theta_{f}^{(1)} n^3}{\Phi_{f}^{(1)}+\Phi_{g}^{(1)}-n (T+2 u)}-\frac{4 b^3 \Theta_{f}^{(1)} n^3}{\Phi_{f}^{(1)}+\Phi_{g}^{(1)}+n (T-2 u)}+\frac{4 b T^2 \Theta_{f}^{(1)} n^3}{\Phi_{f}^{(1)}+\Phi_{g}^{(1)}+n (T-2 u)}
$$
$$
+\frac{4 a b^2 \Phi_{f}^{(2)} \Theta_{f}^{(1)} n^2}{-2 b n+T n-2 u n+\Phi_{f}^{(1)}+\Phi_{g}^{(1)}}+\frac{2 a b^2 (-\Phi_{f}^{(1)}+\Phi_{g}^{(1)}+n T) \Theta_{f}^{(2)} n^2}{-2 b n+T n-2 u n+\Phi_{f}^{(1)}+\Phi_{g}^{(1)}}-\frac{4 a b^2 \Phi_{f}^{(2)} \Theta_{f}^{(1)} n^2}{\Phi_{f}^{(1)}+\Phi_{g}^{(1)}-n (2 b+T+2 u)}
$$
$$
+\frac{2 a b^2 (\Phi_{f}^{(1)}-\Phi_{g}^{(1)}+n T) \Theta_{f}^{(2)} n^2}{\Phi_{f}^{(1)}+\Phi_{g}^{(1)}-n (2 b+T+2 u)}+\frac{2 a b^2 e^{b T} (2 \Phi_{g}^{(2)} \Theta_{f}^{(1)}+(\Phi_{f}^{(1)}-\Phi_{g}^{(1)}-n T) \Theta_{f}^{(2)}) n^2}{\Phi_{f}^{(1)}+\Phi_{g}^{(1)}-n (2 b+T+2 u)}
$$
$$
-\frac{2 a b^2 (2 \Phi_{g}^{(2)} \Theta_{f}^{(1)}+(\Phi_{f}^{(1)}-\Phi_{g}^{(1)}+n T) \Theta_{f}^{(2)}) n^2}{\Phi_{f}^{(1)}+\Phi_{g}^{(1)}+n (T-2 u)}-\frac{1}{-2 b n+T n-2 u n+\Phi_{f}^{(1)}+\Phi_{g}^{(1)}}\bigg[2 a T \Theta_{f}^{(1)} (-\Phi_{f}^{(2)} \Phi_{g}^{(1)}+\Phi_{g}^{(2)} \Phi_{g}^{(1)}
$$
$$
+\Phi_{f}^{(1)} (\Phi_{f}^{(2)}-\Phi_{g}^{(2)})+n \Phi_{f}^{(2)} T+n \Phi_{g}^{(2)} T+2 \Theta_{f}^{(2)} \Theta_{g}^{(1)}+2 \Theta_{f}^{(1)} \Theta_{g}^{(2)}) n\bigg]
-\frac{1}{-2 b n+T n-2 u n+\Phi_{f}^{(1)}+\Phi_{g}^{(1)}}\bigg[2 a b \bigg(-n^2 \Theta_{f}^{(2)} T^2
$$
$$
+n \Phi_{g}^{(2)} \Theta_{f}^{(1)} T-n \Phi_{g}^{(1)} \Theta_{f}^{(2)} T+\Phi_{g}^{(1)} \Phi_{g}^{(2)} \Theta_{f}^{(1)}-\Phi_{f}^{(2)} (\Phi_{g}^{(1)}+n T) \Theta_{f}^{(1)}+\Phi_{f}^{(1)} (\Phi_{f}^{(2)} \Theta_{f}^{(1)}-\Phi_{g}^{(2)} \Theta_{f}^{(1)}+n T \Theta_{f}^{(2)})+2 \Theta_{f}^{(1)} \Theta_{f}^{(2)} \Theta_{g}^{(1)}
$$
$$
+2 \Theta_{f}^{(1)2} \Theta_{g}^{(2)}\bigg) n\bigg]
-\frac{2 a T \Theta_{f}^{(1)} (\Phi_{f}^{(1)} (\Phi_{g}^{(2)}-\Phi_{f}^{(2)})-\Phi_{g}^{(1)} \Phi_{g}^{(2)}+n \Phi_{g}^{(2)} T+\Phi_{f}^{(2)} (\Phi_{g}^{(1)}+n T)-2 \Theta_{f}^{(2)} \Theta_{g}^{(1)}-2 \Theta_{f}^{(1)} \Theta_{g}^{(2)}) n}{\Phi_{f}^{(1)}+\Phi_{g}^{(1)}-n (T+2 u)}
$$
$$
-\frac{1}{\Phi_{f}^{(1)}+\Phi_{g}^{(1)}-n (T+2 u)}\bigg[2 a b \bigg(n^2 \Theta_{f}^{(2)} T^2-n \Phi_{g}^{(2)} \Theta_{f}^{(1)} T+n \Phi_{g}^{(1)} \Theta_{f}^{(2)} T-\Phi_{g}^{(1)} \Phi_{g}^{(2)} \Theta_{f}^{(1)}+\Phi_{f}^{(2)} (\Phi_{g}^{(1)}+n T) \Theta_{f}^{(1)}
$$
$$
-\Phi_{f}^{(1)} (\Phi_{f}^{(2)} \Theta_{f}^{(1)}-\Phi_{g}^{(2)} \Theta_{f}^{(1)}+n T \Theta_{f}^{(2)})-2 \Theta_{f}^{(1)} \Theta_{f}^{(2)} \Theta_{g}^{(1)}-2 \Theta_{f}^{(1)2} \Theta_{g}^{(2)}\bigg) n\bigg]
$$
$$
-\frac{2 a T \Theta_{f}^{(1)} (\Phi_{f}^{(1)} (\Phi_{f}^{(2)}-\Phi_{g}^{(2)})+\Phi_{g}^{(1)} \Phi_{g}^{(2)}-n \Phi_{g}^{(2)} T-\Phi_{f}^{(2)} (\Phi_{g}^{(1)}+n T)+2 \Theta_{f}^{(2)} \Theta_{g}^{(1)}+2 \Theta_{f}^{(1)} \Theta_{g}^{(2)}) n}{\Phi_{f}^{(1)}+\Phi_{g}^{(1)}-n (2 b+T+2 u)}
$$
$$
-\frac{1}{\Phi_{f}^{(1)}+\Phi_{g}^{(1)}-n (2 b+T+2 u)}\bigg[2 a b \bigg(n^2 \Theta_{f}^{(2)} T^2+n \Phi_{g}^{(2)} \Theta_{f}^{(1)} T-n \Phi_{g}^{(1)} \Theta_{f}^{(2)} T-\Phi_{g}^{(1)} \Phi_{g}^{(2)} \Theta_{f}^{(1)}+\Phi_{f}^{(2)} (\Phi_{g}^{(1)}-n T) \Theta_{f}^{(1)}
$$
$$
+\Phi_{f}^{(1)} (-\Phi_{f}^{(2)} \Theta_{f}^{(1)}+\Phi_{g}^{(2)} \Theta_{f}^{(1)}+n T \Theta_{f}^{(2)})-2 \Theta_{f}^{(1)} \Theta_{f}^{(2)} \Theta_{g}^{(1)}-2 \Theta_{f}^{(1)2} \Theta_{g}^{(2)}\bigg) n\bigg]
$$
$$
+\frac{2 a T \Theta_{f}^{(1)} (-\Phi_{f}^{(2)} \Phi_{g}^{(1)}+\Phi_{g}^{(2)} \Phi_{g}^{(1)}+\Phi_{f}^{(1)} (\Phi_{f}^{(2)}-\Phi_{g}^{(2)})+n \Phi_{f}^{(2)} T+n \Phi_{g}^{(2)} T+2 \Theta_{f}^{(2)} \Theta_{g}^{(1)}+2 \Theta_{f}^{(1)} \Theta_{g}^{(2)}) n}{\Phi_{f}^{(1)}+\Phi_{g}^{(1)}+n (T-2 u)}
$$
$$
-\frac{1}{\Phi_{f}^{(1)}+\Phi_{g}^{(1)}+n (T-2 u)}\bigg[2 a b \bigg(-n^2 \Theta_{f}^{(2)} T^2+n \Phi_{f}^{(2)} \Theta_{f}^{(1)} T-n \Phi_{g}^{(2)} \Theta_{f}^{(1)} T+n \Phi_{g}^{(1)} \Theta_{f}^{(2)} T-\Phi_{f}^{(2)} \Phi_{g}^{(1)} \Theta_{f}^{(1)}+\Phi_{g}^{(1)} \Phi_{g}^{(2)} \Theta_{f}^{(1)}
$$
\begin{equation}
+\Phi_{f}^{(1)} (\Phi_{f}^{(2)} \Theta_{f}^{(1)}-\Phi_{g}^{(2)} \Theta_{f}^{(1)}-n T \Theta_{f}^{(2)})+2 \Theta_{f}^{(1)} \Theta_{f}^{(2)} \Theta_{g}^{(1)}+2 \Theta_{f}^{(1)2} \Theta_{g}^{(2)}\bigg) n\bigg]\Bigg]\Bigg/\bigg[2 b n^3 T \left(T^2-b^2\right)\bigg],
\end{equation}
$$
k_{gf}^{(2)}(s,u)=\Bigg[-\frac{4 b \left(b^2-T^2\right) \Theta_{g}^{(1)} n^3}{\Phi_{f}^{(1)}+\Phi_{g}^{(1)}-n (T+2 u)}-\frac{2 a b^2 T \Theta_{g}^{(2)} n^3}{\Phi_{f}^{(1)}+\Phi_{g}^{(1)}-n (2 b+T+2 u)}+\frac{4 b \left(b^2-T^2\right) \Theta_{g}^{(1)} n^3}{\Phi_{f}^{(1)}+\Phi_{g}^{(1)}+n (T-2 u)}
$$
$$
+\frac{2 a b^2 \Phi_{g}^{(1)} \Theta_{g}^{(2)} n^2}{-2 b n+T n-2 u n+\Phi_{f}^{(1)}+\Phi_{g}^{(1)}}-\frac{2 a b^2 (2 \Phi_{g}^{(2)} \Theta_{g}^{(1)}+(\Phi_{f}^{(1)}+n T) \Theta_{g}^{(2)}) n^2}{-2 b n+T n-2 u n+\Phi_{f}^{(1)}+\Phi_{g}^{(1)}}-\frac{2 a b^2 \Phi_{g}^{(1)} \Theta_{g}^{(2)} n^2}{\Phi_{f}^{(1)}+\Phi_{g}^{(1)}-n (T+2 u)}
$$
$$
-\frac{2 a b^2 (2 \Phi_{f}^{(2)} \Theta_{g}^{(1)}-(\Phi_{f}^{(1)}+n T) \Theta_{g}^{(2)}) n^2}{\Phi_{f}^{(1)}+\Phi_{g}^{(1)}-n (T+2 u)}+\frac{4 a b^2 \Phi_{g}^{(1)} \Phi_{g}^{(2)} n^2}{\Phi_{f}^{(1)}+\Phi_{g}^{(1)}-n (2 b+T+2 u)}+\frac{2 a b^2 \Phi_{f}^{(1)} \Theta_{g}^{(2)} n^2}{\Phi_{f}^{(1)}+\Phi_{g}^{(1)}-n (2 b+T+2 u)}
$$
$$
-\frac{2 a b^2 \Phi_{g}^{(1)} \Theta_{g}^{(2)} n^2}{\Phi_{f}^{(1)}+\Phi_{g}^{(1)}-n (2 b+T+2 u)}-\frac{2 a b^2 ((\Phi_{f}^{(1)}-n T) \Theta_{g}^{(2)}-2 \Phi_{f}^{(2)} \Theta_{g}^{(1)}) n^2}{\Phi_{f}^{(1)}+\Phi_{g}^{(1)}+n (T-2 u)}+\frac{2 a b^2 \Phi_{g}^{(1)} \Theta_{g}^{(2)} n^2}{\Phi_{f}^{(1)}+\Phi_{g}^{(1)}+n (T-2 u)}
$$
$$
+\frac{2 a T \Theta_{g}^{(1)} (-\Phi_{f}^{(2)} \Phi_{g}^{(1)}+\Phi_{g}^{(2)} \Phi_{g}^{(1)}+\Phi_{f}^{(1)} (\Phi_{f}^{(2)}-\Phi_{g}^{(2)})+n \Phi_{f}^{(2)} T+n \Phi_{g}^{(2)} T+2 \Theta_{f}^{(2)} \Theta_{g}^{(1)}+2 \Theta_{f}^{(1)} \Theta_{g}^{(2)}) n}{-2 b n+T n-2 u n+\Phi_{f}^{(1)}+\Phi_{g}^{(1)}}
$$
$$
-\frac{1}{-2 b n+T n-2 u n+\Phi_{f}^{(1)}+\Phi_{g}^{(1)}}\bigg[2 a b \bigg(n^2 \Theta_{g}^{(2)} T^2+n \Phi_{g}^{(2)} \Theta_{g}^{(1)} T-n \Phi_{g}^{(1)} \Theta_{g}^{(2)} T-2 \Theta_{f}^{(2)} \Theta_{g}^{(1)2}-\Phi_{g}^{(1)} \Phi_{g}^{(2)} \Theta_{g}^{(1)}
$$
$$
+\Phi_{f}^{(2)} (\Phi_{g}^{(1)}-n T) \Theta_{g}^{(1)}-2 \Theta_{f}^{(1)} \Theta_{g}^{(1)} \Theta_{g}^{(2)}+\Phi_{f}^{(1)} (-\Phi_{f}^{(2)} \Theta_{g}^{(1)}+\Phi_{g}^{(2)} \Theta_{g}^{(1)}+n T \Theta_{g}^{(2)})\bigg) n\bigg]
$$
$$
-\frac{2 a T \Theta_{g}^{(1)} (\Phi_{f}^{(1)} (\Phi_{f}^{(2)}-\Phi_{g}^{(2)})+\Phi_{g}^{(1)} \Phi_{g}^{(2)}-n \Phi_{g}^{(2)} T-\Phi_{f}^{(2)} (\Phi_{g}^{(1)}+n T)+2 \Theta_{f}^{(2)} \Theta_{g}^{(1)}+2 \Theta_{f}^{(1)} \Theta_{g}^{(2)}) n}{\Phi_{f}^{(1)}+\Phi_{g}^{(1)}-n (T+2 u)}
$$
$$
-\frac{1}{\Phi_{f}^{(1)}+\Phi_{g}^{(1)}-n (T+2 u)}\bigg[2 a b \bigg(-n^2 \Theta_{g}^{(2)} T^2+n \Phi_{f}^{(2)} \Theta_{g}^{(1)} T-n \Phi_{g}^{(2)} \Theta_{g}^{(1)} T+n \Phi_{g}^{(1)} \Theta_{g}^{(2)} T+2 \Theta_{f}^{(2)} \Theta_{g}^{(1)2}-\Phi_{f}^{(2)} \Phi_{g}^{(1)} \Theta_{g}^{(1)}
$$
$$
+\Phi_{g}^{(1)} \Phi_{g}^{(2)} \Theta_{g}^{(1)}+2 \Theta_{f}^{(1)} \Theta_{g}^{(1)} \Theta_{g}^{(2)}+\Phi_{f}^{(1)} (\Phi_{f}^{(2)} \Theta_{g}^{(1)}-\Phi_{g}^{(2)} \Theta_{g}^{(1)}-n T \Theta_{g}^{(2)})\bigg) n\bigg]
$$
$$
-\frac{2 a T \Theta_{g}^{(1)} (\Phi_{f}^{(1)} (\Phi_{g}^{(2)}-\Phi_{f}^{(2)})-\Phi_{g}^{(1)} \Phi_{g}^{(2)}+n \Phi_{g}^{(2)} T+\Phi_{f}^{(2)} (\Phi_{g}^{(1)}+n T)-2 \Theta_{f}^{(2)} \Theta_{g}^{(1)}-2 \Theta_{f}^{(1)} \Theta_{g}^{(2)}) n}{\Phi_{f}^{(1)}+\Phi_{g}^{(1)}-n (2 b+T+2 u)}
$$
$$
+\frac{1}{\Phi_{f}^{(1)}+\Phi_{g}^{(1)}-n (2 b+T+2 u)}\bigg[2 a b \bigg(T^2 n^3+\Theta_{g}^{(2)} n^2+T (\Phi_{f}^{(2)} \Theta_{g}^{(1)}-\Phi_{g}^{(2)} \Theta_{g}^{(1)}-\Phi_{f}^{(1)} \Theta_{g}^{(2)}+\Phi_{g}^{(1)} \Theta_{g}^{(2)}) n-\Theta_{g}^{(1)} (-\Phi_{f}^{(2)} \Phi_{g}^{(1)}
$$
$$
+\Phi_{g}^{(2)} \Phi_{g}^{(1)}+\Phi_{f}^{(1)} (\Phi_{f}^{(2)}-\Phi_{g}^{(2)})+2 \Theta_{f}^{(2)} \Theta_{g}^{(1)}+2 \Theta_{f}^{(1)} \Theta_{g}^{(2)})\bigg) n\bigg]
$$
$$
-\frac{2 a T \Theta_{g}^{(1)} (-\Phi_{f}^{(2)} \Phi_{g}^{(1)}+\Phi_{g}^{(2)} \Phi_{g}^{(1)}+\Phi_{f}^{(1)} (\Phi_{f}^{(2)}-\Phi_{g}^{(2)})+n \Phi_{f}^{(2)} T+n \Phi_{g}^{(2)} T+2 \Theta_{f}^{(2)} \Theta_{g}^{(1)}+2 \Theta_{f}^{(1)} \Theta_{g}^{(2)}) n}{\Phi_{f}^{(1)}+\Phi_{g}^{(1)}+n (T-2 u)}
$$
$$
-\frac{1}{\Phi_{f}^{(1)}+\Phi_{g}^{(1)}+n (T-2 u)}\bigg[2 a b \bigg(n^2 \Theta_{g}^{(2)} T^2-n \Phi_{g}^{(2)} \Theta_{g}^{(1)} T+n \Phi_{g}^{(1)} \Theta_{g}^{(2)} T-2 \Theta_{f}^{(2)} \Theta_{g}^{(1)2}-\Phi_{g}^{(1)} \Phi_{g}^{(2)} \Theta_{g}^{(1)}+\Phi_{f}^{(2)} (\Phi_{g}^{(1)}
$$
\begin{equation}
+n T) \Theta_{g}^{(1)}-2 \Theta_{f}^{(1)} \Theta_{g}^{(1)} \Theta_{g}^{(2)}-\Phi_{f}^{(1)} (\Phi_{f}^{(2)} \Theta_{g}^{(1)}-\Phi_{g}^{(2)} \Theta_{g}^{(1)}+n T \Theta_{g}^{(2)})\bigg) n\bigg]\Bigg]\Bigg/\bigg[2 b n^3 T \left(T^2-b^2\right)\bigg],
\end{equation}

$$
k_{gg}^{(2)}(s,u)=\Bigg[\frac{2 b^3 T n^4}{\Phi_{f}^{(1)}+\Phi_{g}^{(1)}-n (T+2 u)}+\frac{2 b^3 T n^4}{\Phi_{f}^{(1)}+\Phi_{g}^{(1)}+n (T-2 u)}+\frac{2 a \Phi_{g}^{(2)} T^3 n^3}{-2 b n+T n-2 u n+\Phi_{f}^{(1)}+\Phi_{g}^{(1)}}
$$
$$
-\frac{2 a b^2 \Phi_{g}^{(2)} T n^3}{-2 b n+T n-2 u n+\Phi_{f}^{(1)}+\Phi_{g}^{(1)}}-\frac{2 a \Phi_{g}^{(2)} T^3 n^3}{\Phi_{f}^{(1)}+\Phi_{g}^{(1)}-n (T+2 u)}+\frac{2 b^3 \Phi_{f}^{(1)} n^3}{\Phi_{f}^{(1)}+\Phi_{g}^{(1)}-n (T+2 u)}-\frac{2 b^3 \Phi_{g}^{(1)} n^3}{\Phi_{f}^{(1)}+\Phi_{g}^{(1)}-n (T+2 u)}
$$
$$
+\frac{2 a b^2 \Phi_{g}^{(2)} T n^3}{\Phi_{f}^{(1)}+\Phi_{g}^{(1)}-n (T+2 u)}-\frac{2 a \Phi_{g}^{(2)} T^3 n^3}{\Phi_{f}^{(1)}+\Phi_{g}^{(1)}+n (T-2 u)}-\frac{2 b^3 \Phi_{f}^{(1)} n^3}{\Phi_{f}^{(1)}+\Phi_{g}^{(1)}+n (T-2 u)}+\frac{2 b^3 \Phi_{g}^{(1)} n^3}{\Phi_{f}^{(1)}+\Phi_{g}^{(1)}+n (T-2 u)}
$$
$$
+\frac{2 a b^2 \Phi_{g}^{(2)} T n^3}{\Phi_{f}^{(1)}+\Phi_{g}^{(1)}+n (T-2 u)}+\frac{2 a b^2 \Phi_{f}^{(1)} \Phi_{g}^{(2)} n^2}{-2 b n+T n-2 u n+\Phi_{f}^{(1)}+\Phi_{g}^{(1)}}+\frac{a b^2 \Phi_{g}^{(2)} T n^2}{u-\frac{-2 b n-T n+\Phi_{f}^{(1)}+\Phi_{g}^{(1)}}{2 n}}-\frac{a \Phi_{g}^{(2)} T^3 n^2}{u-\frac{-2 b n-T n+\Phi_{f}^{(1)}+\Phi_{g}^{(1)}}{2 n}}
$$
$$
-\frac{2 a b^2 \Phi_{g}^{(1)} \Phi_{g}^{(2)} n^2}{-2 b n+T n-2 u n+\Phi_{f}^{(1)}+\Phi_{g}^{(1)}}-\frac{4 a b^2 \Theta_{f}^{(1)} \Theta_{g}^{(2)} n^2}{-2 b n+T n-2 u n+\Phi_{f}^{(1)}+\Phi_{g}^{(1)}}-\frac{2 a T^2 (\Phi_{f}^{(1)} \Phi_{g}^{(2)}-\Phi_{g}^{(1)} \Phi_{g}^{(2)}-\Theta_{f}^{(2)} \Theta_{g}^{(1)}-\Theta_{f}^{(1)} \Theta_{g}^{(2)}) n^2}{-2 b n+T n-2 u n+\Phi_{f}^{(1)}+\Phi_{g}^{(1)}}
$$
$$
+\frac{2 a b^2 \Phi_{f}^{(1)} \Phi_{g}^{(2)} n^2}{\Phi_{f}^{(1)}+\Phi_{g}^{(1)}-n (T+2 u)}-\frac{2 a b^2 \Phi_{g}^{(1)} \Phi_{g}^{(2)} n^2}{\Phi_{f}^{(1)}+\Phi_{g}^{(1)}-n (T+2 u)}-\frac{4 a b^2 \Theta_{f}^{(2)} \Theta_{g}^{(1)} n^2}{\Phi_{f}^{(1)}+\Phi_{g}^{(1)}-n (T+2 u)}
$$
$$
-\frac{2 a T^2 (\Phi_{f}^{(1)} \Phi_{g}^{(2)}-\Phi_{g}^{(1)} \Phi_{g}^{(2)}-\Theta_{f}^{(2)} \Theta_{g}^{(1)}-\Theta_{f}^{(1)} \Theta_{g}^{(2)}) n^2}{\Phi_{f}^{(1)}+\Phi_{g}^{(1)}-n (T+2 u)}-\frac{2 a b^2 \Phi_{f}^{(1)} \Phi_{g}^{(2)} n^2}{\Phi_{f}^{(1)}+\Phi_{g}^{(1)}+n (T-2 u)}+\frac{2 a b^2 \Phi_{g}^{(1)} \Phi_{g}^{(2)} n^2}{\Phi_{f}^{(1)}+\Phi_{g}^{(1)}+n (T-2 u)}
$$
$$
+\frac{4 a b^2 \Theta_{f}^{(2)} \Theta_{g}^{(1)} n^2}{\Phi_{f}^{(1)}+\Phi_{g}^{(1)}+n (T-2 u)}+\frac{2 a T^2 (\Phi_{f}^{(1)} \Phi_{g}^{(2)}-\Phi_{g}^{(1)} \Phi_{g}^{(2)}-\Theta_{f}^{(2)} \Theta_{g}^{(1)}-\Theta_{f}^{(1)} \Theta_{g}^{(2)}) n^2}{\Phi_{f}^{(1)}+\Phi_{g}^{(1)}+n (T-2 u)}+\frac{a b^2 \Phi_{f}^{(1)} \Phi_{g}^{(2)} n}{u-\frac{-2 b n-T n+\Phi_{f}^{(1)}+\Phi_{g}^{(1)}}{2 n}}
$$
$$
+\frac{1}{-2 b n+T n-2 u n+\Phi_{f}^{(1)}+\Phi_{g}^{(1)}}\bigg[2 a b (2 \Phi_{f}^{(2)} \Theta_{f}^{(1)} \Theta_{g}^{(1)}-2 \Phi_{g}^{(2)} \Theta_{f}^{(1)} \Theta_{g}^{(1)}-\Phi_{f}^{(1)} \Theta_{f}^{(2)} \Theta_{g}^{(1)}+\Phi_{g}^{(1)} \Theta_{f}^{(2)} \Theta_{g}^{(1)}+n T \Theta_{f}^{(2)} \Theta_{g}^{(1)}
$$
$$
-\Phi_{f}^{(1)} \Theta_{f}^{(1)} \Theta_{g}^{(2)}+\Phi_{g}^{(1)} \Theta_{f}^{(1)} \Theta_{g}^{(2)}-n T \Theta_{f}^{(1)} \Theta_{g}^{(2)}) n\bigg]
+\frac{1}{-2 b n+T n-2 u n+\Phi_{f}^{(1)}+\Phi_{g}^{(1)}}\bigg[2 a T (2 \Phi_{f}^{(2)} \Theta_{f}^{(1)} \Theta_{g}^{(1)}-2 \Phi_{g}^{(2)} \Theta_{f}^{(1)} \Theta_{g}^{(1)}
$$
$$
-(\Phi_{f}^{(1)}-\Phi_{g}^{(1)}) (\Theta_{f}^{(2)} \Theta_{g}^{(1)}+\Theta_{f}^{(1)} \Theta_{g}^{(2)})) n\bigg]
-\frac{a b^2 \Phi_{g}^{(1)} \Phi_{g}^{(2)} n}{u-\frac{-2 b n-T n+\Phi_{f}^{(1)}+\Phi_{g}^{(1)}}{2 n}}-\frac{2 a b^2 \Theta_{f}^{(1)} \Theta_{g}^{(2)} n}{u-\frac{-2 b n-T n+\Phi_{f}^{(1)}+\Phi_{g}^{(1)}}{2 n}}
$$
$$
-\frac{a T^2 (\Phi_{f}^{(1)} \Phi_{g}^{(2)}-\Phi_{g}^{(1)} \Phi_{g}^{(2)}-\Theta_{f}^{(2)} \Theta_{g}^{(1)}-\Theta_{f}^{(1)} \Theta_{g}^{(2)}) n}{u-\frac{-2 b n-T n+\Phi_{f}^{(1)}+\Phi_{g}^{(1)}}{2 n}}
-\frac{1}{\Phi_{f}^{(1)}+\Phi_{g}^{(1)}-n (T+2 u)}\bigg[2 a T (2 \Phi_{f}^{(2)} \Theta_{f}^{(1)} \Theta_{g}^{(1)}-2 \Phi_{g}^{(2)} \Theta_{f}^{(1)} \Theta_{g}^{(1)}
$$
$$
-(\Phi_{f}^{(1)}-\Phi_{g}^{(1)}) (\Theta_{f}^{(2)} \Theta_{g}^{(1)}+\Theta_{f}^{(1)} \Theta_{g}^{(2)})) n\bigg]
-\frac{1}{\Phi_{f}^{(1)}+\Phi_{g}^{(1)}-n (T+2 u)}\bigg[2 b \bigg(n^3 T^3+n^2 (\Phi_{f}^{(1)}-\Phi_{g}^{(1)}) T^2+a n (T \Theta_{f}^{(2)} \Theta_{g}^{(1)}-\Theta_{f}^{(1)} \Theta_{g}^{(2)})
$$
$$
+a (2 \Phi_{f}^{(2)} \Theta_{f}^{(1)} \Theta_{g}^{(1)}-2 \Phi_{g}^{(2)} \Theta_{f}^{(1)} \Theta_{g}^{(1)}-(\Phi_{f}^{(1)}-\Phi_{g}^{(1)}) (\Theta_{f}^{(2)} \Theta_{g}^{(1)}+\Theta_{f}^{(1)} \Theta_{g}^{(2)}))\bigg) n\bigg]
$$
$$
-\frac{2 a T (2 \Phi_{f}^{(2)} \Theta_{f}^{(1)} \Theta_{g}^{(1)}-2 \Phi_{g}^{(2)} \Theta_{f}^{(1)} \Theta_{g}^{(1)}-(\Phi_{f}^{(1)}-\Phi_{g}^{(1)}) (\Theta_{f}^{(2)} \Theta_{g}^{(1)}+\Theta_{f}^{(1)} \Theta_{g}^{(2)})) n}{\Phi_{f}^{(1)}+\Phi_{g}^{(1)}+n (T-2 u)}
-\frac{1}{\Phi_{f}^{(1)}+\Phi_{g}^{(1)}+n (T-2 u)}
$$
$$
\times\bigg[2 b \bigg(n^3 T^3+n^2 (\Phi_{g}^{(1)}-\Phi_{f}^{(1)}) T^2+a n (\Theta_{f}^{(2)} \Theta_{g}^{(1)}-\Theta_{f}^{(1)} \Theta_{g}^{(2)}) T+a (-2 \Phi_{f}^{(2)} \Theta_{f}^{(1)} \Theta_{g}^{(1)}+2 \Phi_{g}^{(2)} \Theta_{f}^{(1)} \Theta_{g}^{(1)}
$$
$$
+(\Phi_{f}^{(1)}-\Phi_{g}^{(1)}) (\Theta_{f}^{(2)} \Theta_{g}^{(1)}+\Theta_{f}^{(1)} \Theta_{g}^{(2)}))\bigg) n\bigg]
+\frac{1}{u-\frac{-2 b n-T n+\Phi_{f}^{(1)}+\Phi_{g}^{(1)}}{2 n}}\bigg[a b (2 \Phi_{f}^{(2)} \Theta_{f}^{(1)} \Theta_{g}^{(1)}-2 \Phi_{g}^{(2)} \Theta_{f}^{(1)} \Theta_{g}^{(1)}-\Phi_{f}^{(1)} \Theta_{f}^{(2)} \Theta_{g}^{(1)}
$$
$$
+\Phi_{g}^{(1)} \Theta_{f}^{(2)} \Theta_{g}^{(1)}-n T \Theta_{f}^{(2)} \Theta_{g}^{(1)}-\Phi_{f}^{(1)} \Theta_{f}^{(1)} \Theta_{g}^{(2)}+\Phi_{g}^{(1)} \Theta_{f}^{(1)} \Theta_{g}^{(2)}+n T \Theta_{f}^{(1)} \Theta_{g}^{(2)})\bigg]
-\frac{1}{u-\frac{-2 b n-T n+\Phi_{f}^{(1)}+\Phi_{g}^{(1)}}{2 n}}\bigg[a T (2 \Phi_{f}^{(2)} \Theta_{f}^{(1)} \Theta_{g}^{(1)}
$$
\begin{equation}
-2 \Phi_{g}^{(2)} \Theta_{f}^{(1)} \Theta_{g}^{(1)}-(\Phi_{f}^{(1)}-\Phi_{g}^{(1)}) (\Theta_{f}^{(2)} \Theta_{g}^{(1)}+\Theta_{f}^{(1)} \Theta_{g}^{(2)}))\bigg]
\Bigg]\Bigg/\bigg[2 b n^3 T \left(T^2-b^2\right)\bigg],
\end{equation}

\begin{equation}\label{eq:401}
k_{ffns}^{(2)}(s,\tau)= \exp\bigg[\frac{a\Phi_{nsf}^{(2)}(s)}{b}(1+e^{-b\tau/n}) +\Phi_{nsf}^{(1)}(s)\frac{\tau}{n})\bigg],
\end{equation}
where $T=\frac{1}{n}\sqrt{(\Phi_{f}^{(1)}-\Phi_{g}^{(1)})^2+4 \Theta_{f}^{(1)} \Theta_{g}^{(1)}}$, $\psi(s)$ is defined by $\psi(s)=\frac{d}{ds}\Gamma(s)$, $H_{(s)}$ is the $s$-th harmonic number, $C_{F}=\frac{4}{3}$ and $\gamma_{E}$ is the Euler-Lagrange constant. In above equations, the coefficients $\Phi$'s and $\Theta$'s  obtained in Refs. \cite{145,146}. The coefficients $r(s,u)$'s and $q(s,u)$'s in Eqs. (\ref{eq:22}-\ref{eq:25}) are as:
\begin{equation}
r_{11}^{(i)}=k_{gf}^{(i)}(s,u),\quad r_{21}^{(i)}=k_{gg}^{(i)}(s,u),
\end{equation}
$$
r_{21}^{(i)}=k_{gf}^{(i)}(s,u),\quad r_{22}^{(i)}=k_{gg}^{(i)}(s,u),\quad r_{32}^{(i)}=-\frac{2 a_{2} (k_{gf}^{(i)} (s,b_{1}+u)) (k_{gg}^{(i)} (s,b_{1}+u))}{B u}, \quad r_{42}^{(i)}=-\frac{a_{2} (k_{gg}^{(i)} (s,b_{1}+u))^2}{B u},
$$
\begin{equation}
r_{52}^{(i)}=-\frac{a_{2} (k_{gf}^{(i)} (s,b_{1}+u))^2}{B u}.
\end{equation}

$$
q_{11}^{(i)}=k_{ff}^{(i)}(s,u),\quad q_{21}^{(i)}=k_{fg}^{(i)}(s,u),\quad
q_{31}^{(i)}=-\frac{2a_{1}k_{gf}^{(i)}(s,b_{1}+u)k_{gg}^{(i)}(s,b_{1}+u)}{u},\quad
q_{41}^{(i)}=-\frac{a_{1}k_{gf}^{(i)}(s,b_{1}+u)^2}{u},
$$
\begin{equation}
q_{51}^{(i)}=-\frac{a_{1}k_{gg}^{(i)}(s,b_{1}+u)^2}{u}.
\end{equation}

$$
q_{12}^{(i)}=k_{ff}^{(i)}(s,u),\quad q_{22}^{(i)}=k_{fg}^{(i)}(s,u),\quad
q_{32}^{(i)}=-\frac{2a_{1}k_{gf}^{(i)}(s,b_{1}+u)k_{gg}^{(i)}(s,b_{1}+u)}{u},\quad
q_{42}^{(i)}=-\frac{a_{1}k_{gf}^{(i)}(s,b_{1}+u)^2}{u},
$$
$$
q_{52}^{(i)}=-\frac{a_{1}k_{gg}^{(i)}(s,b_{1}+u)^2}{u}, \quad q_{62}^{(i)}=\frac{2 a_{1} a_{2}(k_{gf}^{(i)} (b_{1}+u)) (k_{gg}^{(i)} (2b_{1}+u))^2}{B u (b_{1}+u)}
$$
$$
q_{72}^{(i)}=\frac{4 a_{1} a_{2} (k_{gf}^{(i)} (b_{1}+u)) (k_{gf}^{(i)} (2b_{1}+u)) (k_{gg}^{(i)} (2b_{1}+u))}{B u (b_{1}+u)}+\frac{2 a_{1} a_{2} (k_{gg}^{(i)} (b_{1}+u)) (k_{gg}^{(i)} (2b_{1}+u))^2}{B u (b_{1}+u)},
$$
$$
q_{82}^{(i)}=\frac{4 a_{1} a_{2} (k_{gg}^{(i)} (b_{1}+u)) (k_{gf}^{(i)} (2b_{1}+u)) (k_{gg}^{(i)} (2b_{1}+u))}{B u (b_{1}+u)}+\frac{2 a_{1} a_{2} (k_{gf}^{(i)} (b_{1}+u)) (k_{gf}^{(i)} (2b_{1}+u))^2}{B u (b_{1}+u)}
$$
$$
q_{92}^{(i)}=\frac{2 a_{1} a_{2} (k_{gg}^{(i)} (b_{1}+u)) (k_{gf}^{(i)} (2b_{1}+u))^2}{B u (b_{1}+u)}, \quad q_{102}^{(i)}= -\frac{6 a_{1} a_{2}^2 (k_{gf}^{(i)} (2b_{1}+u))^2 (k_{gg}^{(i)} (2b_{1}+u))^2}{B^2 u (b_{1}+u)^2},
$$
$$
 q_{112}=-\frac{a_{1} a_{2}^2 (k_{gg}^{(i)} (2b_{1}+u))^4}{B^2 u (b_{1}+u)^2}, \quad q_{122}^{(i)}=-\frac{4 a_{1} a_{2}^2 (k_{gf}^{(i)} (2b_{1}+u)) (k_{gg}^{(i)} (2b_{1}+u))^3}{B^2 u (b_{1}+u)^2},  
$$
\begin{equation}
q_{132}^{(i)}=-\frac{4 a_{1} a_{2}^2 (k_{gf}^{(i)} (2b_{1}+u))^3 (k_{gg}^{(i)} (2b_{1}+u))}{B^2 u (b_{1}+u)^2}, \quad q_{142}^{(i)}=-\frac{a_{1} a_{2}^2 (k_{gf}^{(i)} (2b_{1}+u))^4}{B^2 u (b_{1}+u)^2},
\end{equation}
where $B=s\rho^{-s}$. 

The coefficients $c_{k,i}$ ( $k=2,L$ and $i=ns, s, g$), which we used in Eq. (\ref{eq:26}), are as:
 \begin{widetext}

\begin{equation}
c^{(1)}_{ns}=1,\ \ c^{(1)}_{s}=1,\ \ c^{(1)}_{g}=0,
\end{equation}

$$
c^{(2)}_{ns}(s,\tau)=1+C_{F}\frac{\tau}{4\pi}\bigg(-9-\frac{2\pi^2}{3}-\frac{2}{(1+s)^2}+\frac{6}{(1+s)}-\frac{2}{(2+s)^2}+\frac{4}{(2+s)}+3 H_{(s)}+\frac{2 H_{(1+s)}}{1+s}+\frac{2 H_{(2+s)}}{2+s}
$$
$$
+\frac{1}{3}\left(\pi^{2}+6H_{(s)}^{2}-6\psi{'}(1+s)\right)+4\psi{'}(1+s)\bigg),
$$
$$
c^{(2)}_{s}(x,\tau)=c^{(2)}_{ns}(s,\tau),
$$
\begin{equation}
c^{(2)}_{g}(s,\tau)=n_{f} \left(-\frac{2 H_{(1+s)}}{1+s}+\frac{4 H_{(2+s)}}{2+s}-\frac{4 H_{(3+s)}}{3+s}-\frac{2}{1+s}+\frac{2}{(1+s)^2}+\frac{16}{2+s}-\frac{4}{(2+s)^2}-\frac{16}{3+s}+\frac{4}{(3+s)^2}\right).
\end{equation}
\end{widetext}

\end{appendix}

%%%%%%%%%%%%%%%%%%%%%%%%%%%%%%%%%%%%%%%%%%%%%%%%%%%%%%%%%%%%%%

\end{document}